\documentclass[range]{ar-astro2e}%%%Put [range] option here for references citations to
\usepackage{ARAstroBib}
\begin{document}
\input epsf.tex    %<-If you need EPS figures to be
                   %  called in {figure} environment for PC
\input epsf.def   %<-If you need EPS figures to be
                   %  called in {figure} environment for Macintosh
\input psfig.sty

\jname{Annual Reviews of Astronomy and Astrophysics}
\jyear{2014}
\jvol{52}
\ARinfo{1056-8700/97/0610-00}

%\title{Mass Loss from High-Mass Stars}
\title{Mass Loss: Its Effect on the Evolution and Fate of High-Mass Stars}

\markboth{Smith}{Mass Loss}

\author{Nathan Smith \affiliation{Steward Observatory, University of
    Arizona, 933 N.\ Cherry Avenue, Tucsn, AZ 85721, USA; email:
    nathans@as.arizona.edu.}}


\begin{abstract}

  Our understanding of massive star evolution is in flux, due to
  recent upheavals in our view of mass loss, and observations of a
  high binary fraction among O-type stars.  Mass-loss rates for
  standard metallicity-dependent line-driven winds of hot stars are
  now thought to be lower by a factor of 2-3 compared to rates adopted
  in modern stellar evolution models, due to the influence of clumping
  on observed diagnostics.  Weaker line-driven winds shift the burden
  of H-envelope removal elsewhere, so that the dominant modes of mass
  loss are the winds, pulsations, and eruptions of evolved
  supergiants, as well as binary mass transfer.  Studies of
  stripped-envelope supernovae, in particular, require binary mass
  transfer.  Dramatic examples of eruptive mass loss are seen in Type
  IIn supernovae, which have massive shells ejected just a few years
  before core collapse. These are a prelude to core collapse, and may
  signify severe instabilities in the latest nuclear burning phases.
  The shifting emphasis from steady winds to episodic mass loss is a
  major change for low-metallicity regions, since eruptions and binary
  mass transfer are less sensitive to metallicity.  We encounter the
  predicament that the most important modes of mass loss are also the
  most uncertain, undermining the predictive power of single-star
  evolution models beyond core H burning.  Moreover, the influence of
  winds and rotation in models has been evaluated by testing
  single-star models against observed statistics that, as it turns
  out, are heavily influenced by binary evolution.  Altogether, this
  alters our view about the most basic outcomes of massive-star mass
  loss --- are Wolf-Rayet stars and Type Ibc supernovae the products
  of single-star winds, or are they mostly the result of binary
  evolution and eruptive mass loss?  This is not fully settled, but
  mounting evidence points toward the latter.  This paradigm shift has
  far-reaching impact on other areas of astronomy, since it changes
  predictions for ionizing radiation and wind feedback from stellar
  populations, it may alter conclusions about star formation rates and
  initial mass functions in external galaxies, it affects the origin
  of various compact stellar remnants, and it determines how we use
  supernovae as probes of stellar evolution across cosmic time.

\end{abstract}

\maketitle

\tableofcontents

\section{INTRODUCTION}

Because of the scaling of luminosity with initial mass, relatively
rare stars born with masses above $\sim$10-20 $M_{\odot}$ vastly
outshine the much larger number of lower-mass stars.  The radiative
output of massive stars is so intense that photon momentum can drive
strong winds, and the energy transport through the stellar interior
may jeaprodize the stability of the star itself.  Mass loss from
massive stars has a {\it deterministic} influence on the structure and
evolution of those stars, which in turn have tremendous impact on
other areas of astronomy.

Massive stars are the cosmic engines that provide most of the
luminosity in star-forming galaxies.  Since massive stars have short
lifetimes, their ultraviolet (UV) radiation (reprocessed in various
forms) is the most observable tracer of star formation and is used to
calculate the star formation rates \citep{k98,ke12}.  Feedback in the
form of UV radiation, stellar winds, and supernovae (SNe) stirs
interstellar gas, driving turbulence and perhaps triggering new
generations of stars by sweeping gas into dense filaments
\citep{el77}.  Through this feedback, massive stars have a profound
impact on the evolution of disk galaxies \citep{vdkf11}.  Massive star
feedback may also terminate star formation locally, blowing giant
bubbles \citep{c05} and leaking hot processed gas into the Galactic
Halo \citep{putman12}.  Elemental yields from nuclear burning in the
cores of massive stars and explosive nucleosynthesis in SNe provide
the elements in the periodic table that pollute the interstellar
medium (ISM), driving galactic chemical evolution and the metallicity
($Z$) evolution of the Universe.  Through their violent deaths as SNe,
massive stars provide brilliant displays that permit us to dissect
individual stellar interiors at distances of many Mpc while the star's
inner layers are peeled away for us to see \citep{filip97}.  They
leave behind exotic corpses such as black holes, neutron stars,
pulsars, magnetars, and all the bizarre high-energy phenomena that
occur when compact objects remain bound in a binary system (e.g.,
\citealt{rm06}).  Shock fronts in their beautifully complex SN
remnants echo through the ISM for thousands of years after the bright
SN display has faded \citep{r08}.

Mass loss affects a star's luminosity, burning lifetime, apparent
temperature, the hardness of its emitted radiation field, its He core
mass, and it will profoundly impact the end fate of a star.  As a
consequence, changes in estimates of mass-loss rates can alter
expectations for their collective ionizing radiation, UV luminosity,
winds, and SNe.  Thus, understanding massive stars and their mass loss
also remains important as we push ever farther to the distant reaches
of the Universe.  Inferences about the initial mass function (IMF;
\citealt{b10}) and variation of the star-formation rate through cosmic
time \citep{madau98,hb06} hinge upon converting reprocessed massive
star UV luminosity to a collective star-formation rate.
%SN explosions are also used as a proxies for star-formation rates.
%Since both of these conversions depend on our understanding of mass
%loss, which is poorly constrained, disagreement between these two
%tracers should not be suprising. 
Gamma ray bursts (GRBs; \citealt{g09,wb06}) and SNe also provide a
probe of very distant populations and stellar evolution in early
environments, but this requires us to understand the connection
between stars of various types, their mass loss, and the eventual type
of SN seen.  The first stars in $Z$-poor environments are expected to
be very massive \citep{bl04}, and hence, the scaling of mass loss with
$Z$ affects abundances of the lowest $Z$ stars \citep{bc05,by11}
polluted by a small number of early SNe.  Since dense stellar winds
can absorb portions of the Lyman continuum emitted by a hot star's
photosphere \citep{paco96}, estimates of mass-loss rates and their
scaling with metallicity can profoundly impact topics as remote as
reionization of the universe and the interpretation of spectra from
galaxies at the highest redshifts currently being detected
\citep{lb01,f06,mw10}.

Although convenient recepies and simple scaling relations to account
for the collective effects of mass loss and feedback will always be
available, researchers working in these other branches of astronomy
should not believe that such recepies are reliable to better than
order-of-magnitude levels.  Even in the very local universe where we
have excellent multiwavelength observations, there is still tremendous
uncertainty in derived mass-loss rates for massive stars, and so there
is large uncertainty in its influence on evolution.  This is
exacerbated by the predicament that these uncertainties are largest
for the most massive and most luminous stars, but these stars also
tend to be the most influential.  Extrapolating to the early Universe
is still quite risky, and should always raise eyebrows.  The aim of
this review article is to provide a broad overview of the current
understanding of mass loss and its influence on the evolution of
massive stars, and to raise a flag of caution about the uncertainties
involved.

\subsection{The Importance of Mass Loss in Massive-Star Evolution}

For low- and intermediate-mass stars ($M_{ZAMS} < 8 M_{\odot}$), wind
mass loss is relatively unimportant for evolution until the final
stages as the asymptotic giant branch (AGB) star transitions to a
protoplanetary nebula.  For massive stars, however, mass loss cannot
be ignored.  For most of their lives --- even on the main sequence ---
massive stars above $\sim$20 $M_{\odot}$ shed mass in fast winds that
affect their subsequent evolution (i.e. $\int\dot{M}dt$ can be a
significant fraction of the stellar mass), and in post-main sequence
phases the mass loss becomes critical in determining the type of
resulting SN explosion.

Thus, mass loss is inexorably linked to evolution for massive stars.
This article will not provide a detailed review of our understanding
of stellar evolution from a theoretical perspective, since this has
already been done in a number of excellent reviews. Although somewhat
dated, \citet{cm86} provide a good description of how models that
incorporate mass loss compare to stellar evolution models without mass
loss.  \citet{mm00} provide a review of how the inclusion of rotation
can influence single-star stellar evolution models, while
\citet{langer12} has reviewed more recent advances as well as
important aspects of how close binarity may dramatically change the
evolutionary paths of stars.  Finally, \citet{whw02} discuss stellar
evolution models with particular emphasis on the late pre-SN burning
phases and their connection to core collapse.

Instead, this review will concentrate mostly on observational
estimates of mass-loss and their impact on stellar evolution, because
this is where the largest uncertainty currently resides.  Stellar
evolution calculations must adopt prescriptions for mass-loss rates as
input to their code, but these assumed rates determine the outcome of
evolution.  Depending on precarious assumptions about wind strength, a
red supergiant (RSG) can be made to evolve to the blue on the
Hertzsprung-Russell (HR) Diagram, or not.  A massive star can be
driven to the luminous blue variable (LBV) phase, or it can avoid it
altogether. Moreover, the mass-loss rates typically used in stellar
evolution models are time-averaged, even though eruptive and explosive
mass loss events are observed.  Hence, there is great uncertainty in
the predictions of all evolutionary models.  Tracks on the HR diagram
are not plotted with error bars that reflect these uncertain
assumptions.

Due to the need for time-averaged prescriptions in stellar evolution
calculations, the steady line-driven winds of hot stars have consumed
the majority of effort in understanding mass loss in the massive star
community for the past three decades, both theoretically and
observationally. There have been great leaps in quantitative non-LTE
modeling of spectra influenced by winds.  A previous review by
\citet{kp00} (KP00 hereafter) dealt with line-driven winds of hot
stars, focussing on the theory of the driving mechanism as well as
common scaling relations used to convert observables to rates.  A more
recent review by \citet{puls08} also concentrated on the relatively
steady line-driven winds of hot stars, providing a thorough discussion
of wind theory and its connection to a wide array observational
diagnostics.  Those reviews did not focus on more extreme winds of
evolved massive stars or mass loss in binaries, and they did not
discuss highly time-dependent mass loss associated with transient
events (eruptions, explosions, mergers).  This review will therefore
emphasize these latter topics, but it will discuss more recent
developments, in particular the reduction in mass-loss rates and the
consequent paradigm shift in stellar evolution.

\subsection{Historical Perspective and Paradigm Shift}

Our present understanding of mass loss has undergone dramatic changes,
due in large part to major shifts in our quantitative estimates of
what mass-loss rates actually are.  The historical path to our current
understanding has had some interesting swings.

Very early observations of transient events like Tycho's SN, P Cygni's
1600 AD eruption, and many nova eruptions, as well as the broad
emission lines in Wolf-Rayet (WR) stars \citep{wr1867} and the broad
blueshifted absorption seen in spectra of many objects indicated the
presence of outflowing material.  However, these were seen as rare
stars or brief eruptive events, and the connection to the lives of
normal stars was unclear.  Later, estimates for the solar wind
\citep{parker58} powered by hot gas pressure had such low mass-loss
rates that winds would not matter much in stellar evolution.  There
were some important early indications and expectations of mass loss
from hot stars \citep{sobolev60}, but the mass-loss rates were very
uncertain.  Therefore, the dominant paradigm (e.g.,
\citealt{pac66,pac67,pac71}) was that, as in low-mass stars, binary
Roche-Lobe Overflow (RLOF) played a major role in making massive
stripped-envelope stars like the He-rich WR stars.

The birth of UV astronomy triggered a revolution in our understanding
of massive stars by providing the decisive evidence that all luminous
hot stars have strong winds, indicated by deep P~Cyg absorption in
their UV resonance lines \citep{morton67}.  This led to a new paradigm
wherein line-driven winds dominate the stripping of the H envelope,
rather than binaries, leading to WR stars and Type Ibc SN progenitors
via single-star evolution (i.e., the so-called ``Conti scenario'';
\citealt{conti76}).  As with UV observations, the development of IR
detectors led to a similar recognition of the importance of
dust-driven winds in red supergiants \citep{gw71}.  With the
assumption that mass loss in steady winds dominates mass loss,
theorists could adopt simple prescriptions for that mass loss and
calculate single-star evolutionary tracks on the HR Diagram (see
reviews by \citealt{cm86,mm00}).  These single-star models were able
to provide a plausible explanation for observed distributions of
stars, including the relative amounts of time spent in different
evolutionary phases as O-type stars, WR stars, and RSGs.
\citet{massey03} has reviewed how these single-star models have been
compared to observations and how observed statistics have been used to
test and refine the single-star models.  This single-star paradigm
then permitted one to extend even further, to compute models for the
collective radiative output of entire stellar populations as a
function of age (e.g., codes like {\sc starburst99};
\citealt{leitherer99}).

%Looking at conference proceedings from 1980s and 1990s, much of the
%community forgot about things called bianries - with rare exceptions
%about free parameters.

%%%%%%%%%%%

In the last decade this paradigm has shifted yet again because of two
important realizations: 1) due to the effects of clumping (see below),
empirical mass-loss rates for line-driven winds are lower than
previously thought, and 2) other unsteady modes of mass loss are more
important than previously recognized, due to increased estimates for
mass ejected in episodic mass loss events \citep{so06} and the very
high binary fraction among massive stars \citep{sana12}.  These
changes to our view of mass loss are discussed in detail below. This
may alter a huge number of predictions for the evolutionary tracks of
stars and their variation with $Z$.  The pendulum appears to be
swinging back to binaries as the dominant agent in massive star
evolution, although not everyone is on the bandwagon.  Justifiably,
this topic occupies much of the current debate and excitement in the
field.

\section{THE DIMINISHED ROLE OF STEADY LINE-DRIVEN WINDS}

Massive stars spend the majority of their lives as hot stars (mostly
as OB types, and more briefly as WR stars), and during these phases
the dominant mode of mass loss is through line-driven winds.  Again,
see KP0 and \citet{puls08} for reviews concerning steady, line-driven
winds from hot stars.  Here we focus on the treatment of winds in
evolution, as well as important updates.

\begin{figure}%3	% Figure using psfig.sty
%\centerline{\psfig{figure=hegerFig2.eps,height=20pc}}
%\centerline{\psfig{figure=hegerMZ-SN.eps,height=25pc}}
\centerline{\psfig{figure=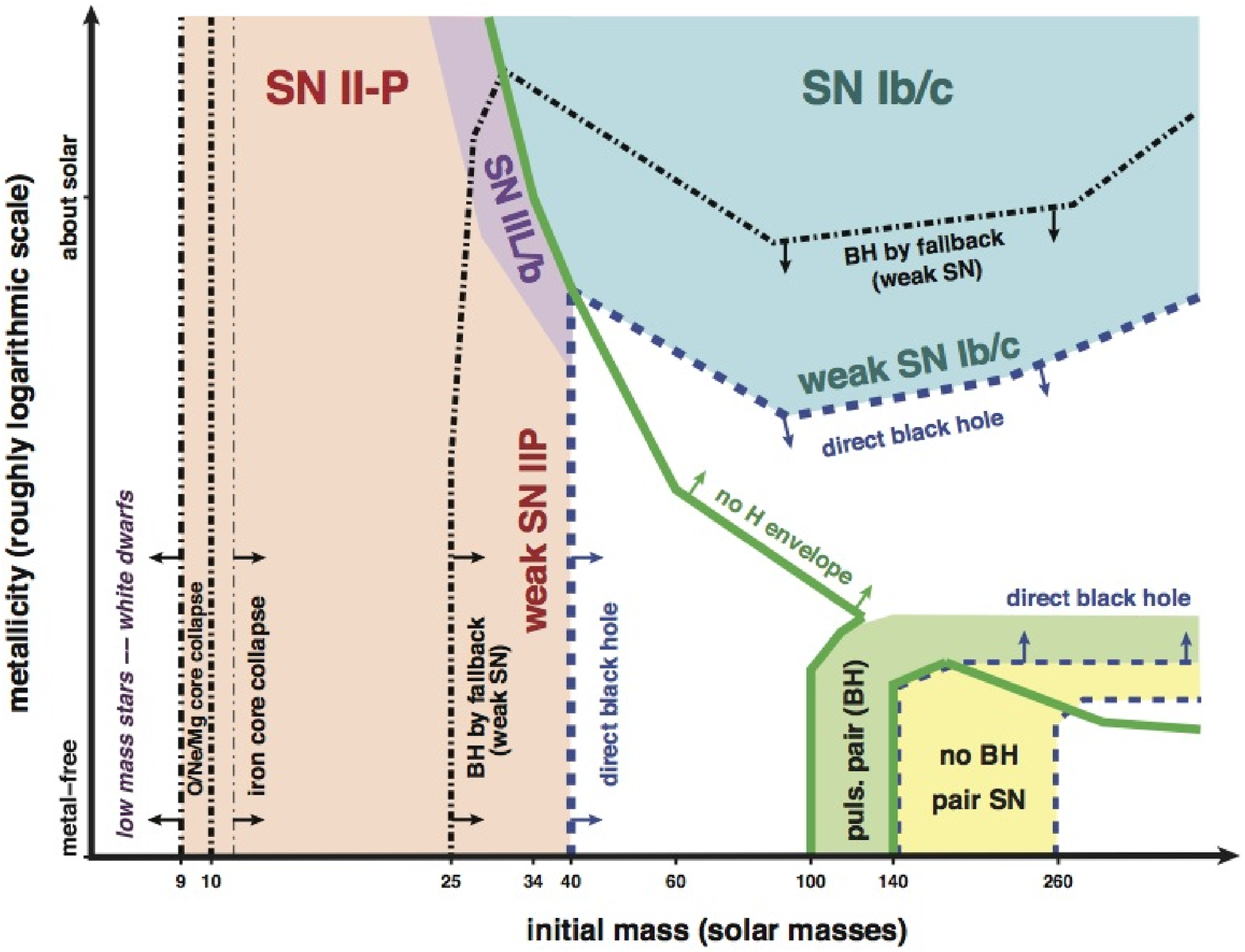,height=25pc}}
\caption{Example of the standard expectations for the fates of massive
  stars as a function of initial mass and metallicity, adapted from
  \citet{heger03} with permission.}
\label{fig:heger}
\end{figure}

\subsection{The Standard View}

In the wind of a hot star, momentum is transfered from the outwardly
propagating radiation to the gas through absorption and scattering by
UV metal lines \citep{ls70,cak75}.  Thus, the rate at which mass is
lifted from the star by this mechanism depends on the UV luminosity of
the star, the temperature (and ionization), and the metallicity, $Z$
\citep{puls08,mokiem07}.  Most stellar evolution models therefore
adopt simple prescriptions for the mass-loss rates that scale smoothly
with the stellar luminosity, temperature, and metallicity.  For the
generation of models calculated throughout the 1990s, the most
commonly adopted mass-loss rates were those of \citet{dj88} and
\citet{ndj90} for O-type stars and RSGs, and \citet{nl00} for WR
stars.  A representative example of the application of this technique
to infer the evolution and fate of massive stars as a function of
metallicity is discussed by \citet{heger03}, where $Z$-dependent mass
loss ($\propto \sqrt{Z}$) is adopted (many other stellar evolution
models are reviewed by \citealt{langer12}).  Hence, it is expected
that the total mass lost by a star will increase smoothly with
luminosity and $Z$, yielding trends with $M_{ZAMS}$ and $Z$ such as
that shown in Figure~\ref{fig:heger} \citep{heger03}.  This basic
picture has been widely regarded as the ``standard view'' of
single-star evolution at high initial mass, although details of the
implementation differ from one model to the next.

\subsection{Observational Diagnostics and Clumping}

The mass-loss rate, $\dot{M} = 4 \pi r^2 \rho(r) v_{\infty}$, depends on
the average mass density $\rho(r)$ at a particular radius where the
wind has reached its terminal velocity $v_{\infty}$.  Connecting these
ideal values to observations is non-trivial and requires detailed
models including realistic opacities, since the radius where the
emissivity or absorption originates is wavelength dependent.  The wind
terminal speed can be deduced from resolved profiles of P Cygni
absorption in UV resonance lines.  There are several diagnostics of
the wind density, the most common being the strength of wind free-free
emission in the IR or radio \citep{wb75}, H$\alpha$ or other
recombination emission lines, and the strength of blueshifted P Cygni
absorption features in unsaturated UV resonance lines (see
\citealt{puls08}).

Radio/IR free-free continuum excess and emission lines like H$\alpha$,
He~{\sc i}, and He~{\sc ii} are recombination processes, so their
emissivity varies as $\rho^2$, whereas P~Cyg absorption varies
linearly with $\rho$.  The quadratic density dependence of
recombination emissivity raises a problem --- if small-scale
inhomogeneities (i.e. ``clumps'') permeate the wind, then
recombination emission arising in dense clumps will be stronger than
emission from the same amount of mass distributed uniformly throughout
the wind (in other words, $\langle{\rho^2}\rangle >
\langle{\rho}\rangle^2$).  It is now well-established that winds are
in fact clumpy (see below), so when mass-loss rates are derived from
H$\alpha$ or free-free excess using the assumption of a homogeneous
wind, {\it the mass-loss rates are overestimated}.  This is the case
for the frequently used ``standard'' mass-loss rates of \citet{dj88}
and \citet{ndj90}, which assumed a smooth wind.  The factor by which
mass-loss rates are overestimated is $\sqrt{f_{cl}}$, where it is
standard practice to define the ``clumping factor'' as $f_{cl} =
\langle{\rho^2}\rangle / \langle{\rho}\rangle^2$. (This assumes that
the gas is optically thin.)  Constraining the value of $f_{cl}$
observationally is paramount for understanding stellar evolution.

Significant wind clumping is expected on theoretical grounds
\citep{or84,owocki88,feld95,op99,do05,so13}, due mostly to the
line-driven instability (this arises because the force of line driving
is velocity dependent; gas parcels which absorb line photons are
accelerated, and are therefore Doppler shifted out of the line to
absorb adjacent photons).  Clumps are expected on a size scale
comparable to the Sobolov length, given by the thermal velocity
divided by the radial velocity gradient in the wind ($dv/dr$), which
means that clumping should exist on a size scale smaller than the
stellar radius.  Clumping may also be induced at the base of the wind
because of sub-surface convection driven by the Fe opacity bump
\citep{cantiello09}.  In fact, it has long been known that hot-star
winds are probably clumpy on small scales
\citep{hillier91,mr94,drew94} and inhomogeneous on larger scales;
large scales include such complexities as the time-variable discrete
absorption components \citep{howarth95,massa95,co96,fullerton97}, or
axisymmetric winds due to rapid rotation \citep{owocki96}, and a wide
array of magnetically induced inhomogeneities \citep{uo02,townsend05}.
However, it was only recently that the severity of the problem for the
global mass-loss rate was quantified.

\begin{figure}%3	% Figure using psfig.sty
%\centerline{\psfig{figure=fullertonFig.eps,height=20pc}}
\centerline{\psfig{figure=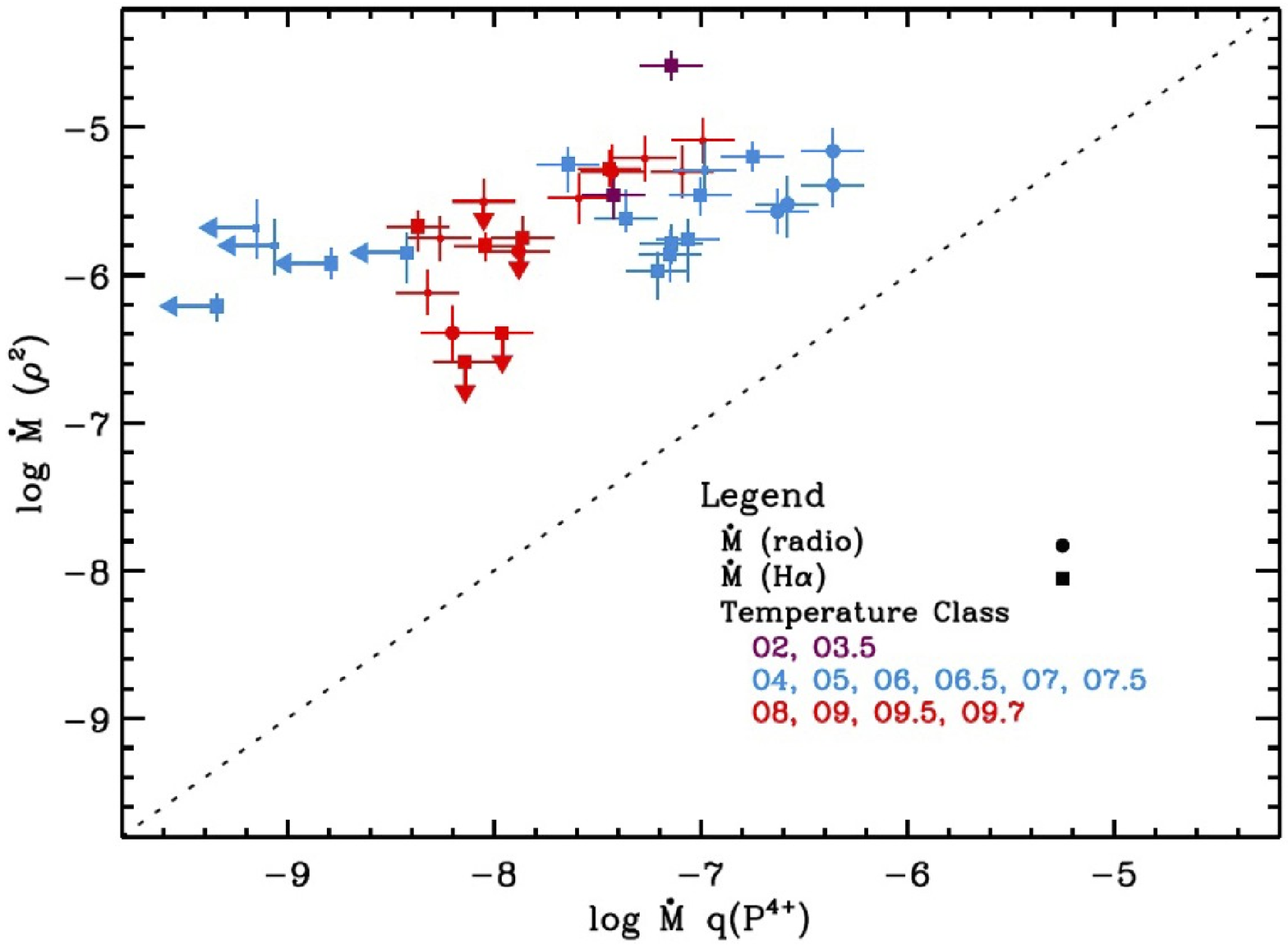,height=20pc}}
\caption{A comparison of the mass-loss rates derived from diagnostics
  that are linearly proportional to wind density, like UV P~Cyg
  absoriton, and those which are proportional to $\rho^2$, like
  free-free and H$\alpha$ emission.  This is from \citet{fullerton06},
  reproduced with permission.  Although there is still discussion
  about the P~{\sc v} lines used for this study and if they may
  overestimate the reductions in $\dot{M}$ that are applied, this
  nevertheless forced an important discussion in massive star research
  by highlighting the potential influence of clumping.}
\label{fig:fullerton}
\end{figure}

One can check the influence of clumping on $\rho^2$ diagnostics (and
thereby measure $f_{cl}$) by studying the same winds using diagnostics
that are linearly proportional to density, like P Cyg absorption in UV
resonance lines and the strengths of electron-scattering wings.  Using
UV resonance absorption lines in O-type stars, \citet{fullerton06}
have proposed a reduction by a factor of 10 or more from traditional
mass-loss rates ($f_{cl}$ values of 100 or more), while
\citet{bouret05} require reductions by factors of $\sim$3.  For the
Milky Way, LMC, and SMC, various studies using modern non-LTE analysis
find $f_{cl} \simeq 10$
(\citealt{puls06,crowther02,figer02,hillier03,massa03,evans04}),
corresponding to $\dot{M}$ reductions by a factor of $\sim$3 with
typical uncertainties of $\sim$30\%.  Based on unifying H$\alpha$
meaurements with the theoretical mind momentum relation, both
\citet{repolust04} and \citet{markova04} found $f_{cl}$=5, or a
mass-loss raduction by 2.3.\footnote{Note that long ago, small scale
  clumping with $f_{cl}$ values of 4-20 (a reduction in mass-loss
  rates by factors of 2-4) was required to fit both the emisson cores
  ($\propto \rho^2$) and electron-scattering wings ($\propto \rho$) in
  WR stars \citep{hillier91,mr94}.  Line wings of O-type stars are too
  weak for this analysis.  In addition, polarization variations in
  WR+O eclipsing binaries \citet{stlouis93} implied $f_{cl} \simeq
  10$.}

\citet{puls06,puls08} and others have discussed that the larger
mass-loss rate reduction of 10 found by \citet{fullerton06} may be an
overestimate because of how ionization can affect the optically thin
P~{\sc v} lines that the large clumping factor is based upon, as well
as possible mediating effects of porosity in the wind.  These topics
are not completely settled,\footnote{The interested reader can consult
  the proceedings volume of a recent conference on this topic for more
  detailed information \citep{hfo08}.} but nevertheless, most
observational studies agree that for mid/early O-type stars, clumping
is significant enough to warrant mass-loss rates reduced by a factor
of 2 to 3 relative to the standard rates from H$\alpha$ and radio flux
that assume homogeneous winds \citep{dj88,ndj90}.  Reductions by
factors of 2-4 are confirmed by X-ray observations
\citep{kramer03,cohen10,cohen11}.  Thus, reductions in mass-loss rates
for normal O-type stars by a factor of 3 ($\pm$30\%) is a good guide.

\begin{figure}%3	% Figure using psfig.sty
%\centerline{\psfig{figure=../Dropbox/massloss.eps,height=30pc}}
\centerline{\psfig{figure=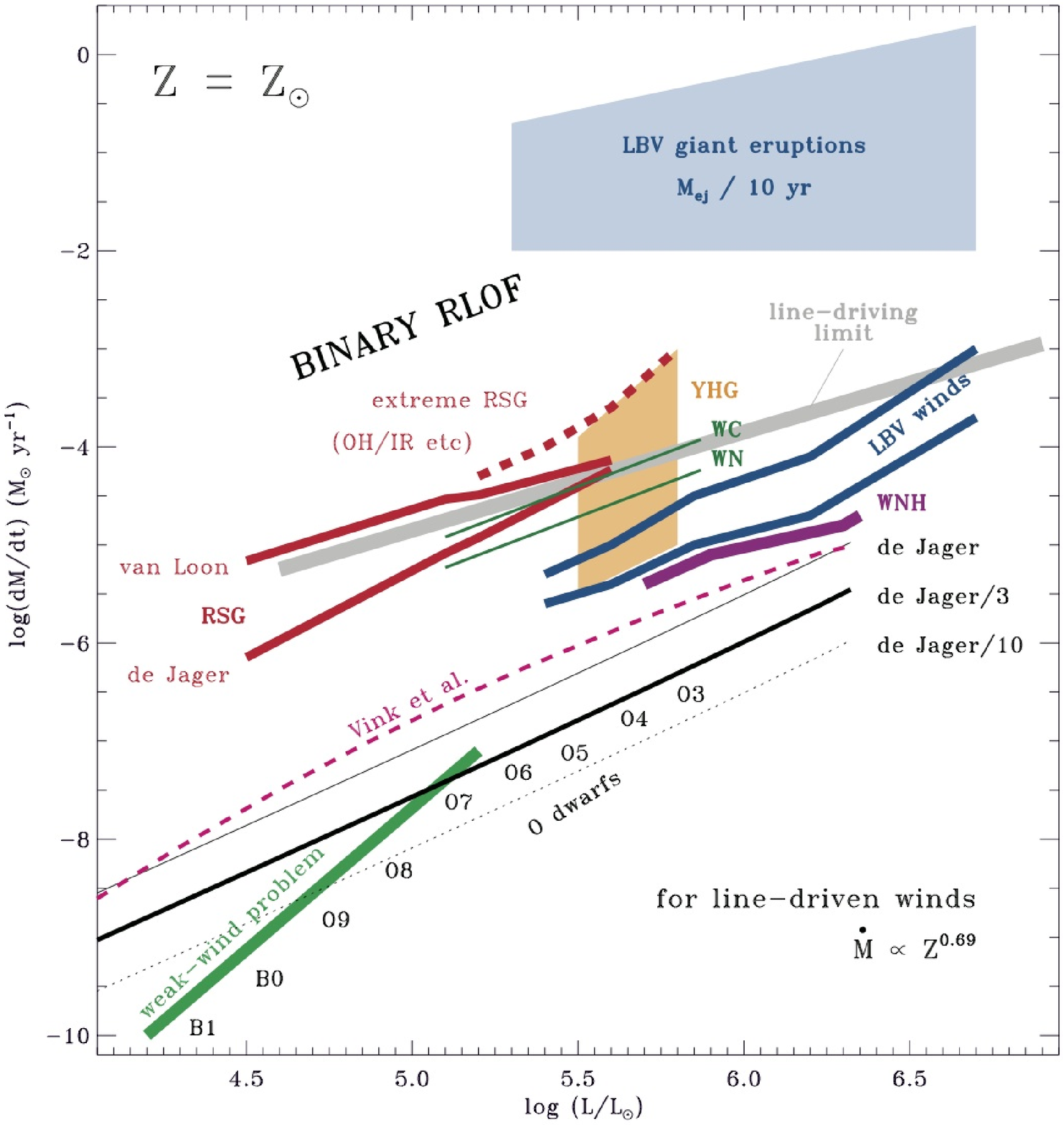,height=30pc}}
\caption{\footnotesize A number of different prescriptions for wind
  mass loss used in models, as well as typical observed ranges of
  mass-loss rates for a number of different types of stars.  For
  O-type stars, the theoretical rates from the prescription of
  \citet{vink01} are shown, along with ``standard'' observational
  rates using the prescription from \citet{dj88}, as well as these
  same prescriptions divided by factors of 3 and 10 for comparison.
  The green line labeled ``weak-wind problem'' refers to lower
  mass-loss rates for late O-type and early B-type MS stars (see
  text).  Rates for WN and WC stars are from \citet{crowther07}. RSG
  mass-loss prescriptions are from \citet{dj88} and \citet{vanloon05},
  as indicated. For YHGs, see \citet{dj98}.  For $\dot{M}$
  corresponding to normal winds of LBVs, values were compiled from a
  number of studies \citep{vdk02,svdk04,groh09,hillier01}.  For LBV
  eruptions, the ``rates'' shown are calculated from total masses
  observed in LBV CSM shells \citep{so06} divided by a nominal
  eruption duration of 10 yr (see Figure~5). For ``binary RLOF'', an
  order-of-magnitude value for the strongest mass-transfer rates
  expected in brief RLOF phases is noted, although the mass-transfer
  or mass-loss rate can be much less for slow mass transfer, or
  possibly more for dynamical common-envelope ejection events; see
  references in the text, especially the review by \citet{langer12}.}
\label{fig:mdot}
\end{figure}

\subsection{Recent Developments and Modifications}

In current generations of stellar evolution models (see the recent
review by \citealt{langer12}), the most commonly used prescription for
the mass-loss rates of hot stars with line-driven winds is from
\citet{vink01}.  These are theoretical mass-loss rates based on the
expected radiative acceleration of a wind, calculated by loss of
photon energy using a Monte-Carlo method.  These $\dot{M}$ values are
comparable to the values of the old ``standard'' rates derived
observationally from $\rho^2$ diagnostics assuming homogeneous winds.
The prescriptions for O-star mass-loss rates taken directly from
\citet{dj88} and \citet{vink01} are compared in Figure~\ref{fig:mdot}
(this figure also includes $\dot{M}$ for other types of stars
discussed in this review).  For comparison, Figure~\ref{fig:mdot} also
plots the \citet{dj88} rates divided by factors of 3 (the favored
reduction) and 10 (possibly an overestimate) to account for the way
clumping affects the observationally derived rates.  These $\dot{M}$
prescriptions are for O-type MS stars over a range of luminosities
(using different $M_{ZAMS}$ and $T_{eff}$ values from stellar
evolution models) at $Z_{\odot}$.  All these values of $\dot{M}$
increase with both increasing $L$ and decreasing $T_{eff}$, so
$\dot{M}$ will climb by a factor of 3--4 as an O-type star evolves
along the MS, giving rise to observed properties Of and WNH stars.
The most luminous WNH stars are generally assumed to be in the late
phases of core H burning, rather than He burning like traditional WR
stars, but some other WN stars with H in their spectra are indeed
thought to be transition objects and possibly related to LBVs.  Binary
evolution may of course play a role in creating some of these WN stars
with H, and the connections between terminology and evolutionary state
are often complicated.

An important point to recognize is that the theoretical \citet{vink01}
prescription for $\dot{M}$ is almost the same as the empirical
\citet{dj88} prescription (in fact, for much of the range of O-star
luminosities, the Vink et al.\ mass-loss rates are higher).  This is
for physical parameters from model ZAMS stars from \citet{ekstrom12};
note that the Vink et al. recipe depends on $L$, $_{\rm eff}$, $M$,
and $v_{\infty}$, whereas the de Jager prescription uses only $L$ and
$T_{\rm eff}$.  The two prescriptions therefore change differently as
a star evolves.  In any case, it appears that stellar evolution
calculations {\it are still using mass-loss rates that are too high}
by a factor of $\sim$3 during the MS lifetimes of massive
stars.\footnote{The \citet{vink01} mass-loss rates are indeed a factor
  of $\sim$2 lower than rates used in {\it some} massive star
  evolution models. In particular, older models by \citet{meynet94}
  adopted mass-loss rates that were artificially a factor of 2 higher
  than the \citet{dj88} mass-loss rates, because this enhanced mass
  loss did a better job of accounting for the observed statistics of
  WR stars.  Moreover, these models with enhanced mass-loss rates are
  still often employed in stellar population synthesis models.}  It is
likely that adopting the reduced mass-loss rates will have a profound
impact on the outcome of single-star evolution calculations, but a
meaningful comparison with observed properties of stars cannot be
attempted until this is updated. 

Modifications to standard mass-loss rates have also occured at the
high and low ends of the luminosity range.  As the most massive O-type
stars (spectral types of O3 and O2) evolve toward the terminal-age MS,
their luminosities go up and they move close to $\Gamma$=1, where
$\Gamma = \kappa L / 4 \pi GM c$ is the Eddington ratio.  High
$\Gamma$ values in hot stars can substantially affect the winds and
increase mass-loss rates \citep{grafener11,vink11}.  Thus, it may be
possible for stars with initial masses of 80-100 $M_{\odot}$ or more
to lose mass so fast in a WNH phase that they avoid the LBV phase
altogether, although this effect has not yet been included in stellar
evolution calculations.  For these very massive stars it is likely
that steady winds could have a more significant impact on evolution
than for the majority of SN progenitors with lower initial masses.
%Moreover, it is unclear if there is any observational evidence for the
%very massive H-poor WR stars that should result from this track, and
%SN Ibc ejecta masses seem to challenge this (see below).  Of courwe,
%such winds may also apply to the most massive stars that result from
%binary mergers or mass accretion.

While the most massive H-burning stars may have winds that are
stronger than the standard \citet{vink01} prescriptions, it has been
found that later O-type and early B-type stars have surprisingly weak
winds for their luminosity compared to theoretical expectations.
Below log($L/L_{\odot}$)=5.2 (spectral types of O7 and later),
observed wind momenta and mass-loss rates are much lower than
theoretical predictions (see Figure~\ref{fig:mdot}).  This is known as
the ``weak-wind problem'' for O dwarfs \citep{puls08,muijres12}, and
may indicate inefficient line driving in hot dwarfs.  Analysis of the
bow shock around the O9.5 V runaway star $\zeta$ Oph suggests that the
weak-wind problem may not be as bad as indicated by UV absorption
(factor of 100 lower), but that the mass-loss rates are still a factor
of 6--7 lower \citep{gvaramadze12}.  Similarly, \citet{huen12} find
from considering X-ray diagnostics that the $\dot{M}$ reduction may
not be as severe as suggested from UV diagnostics, although their
favored rate of $\dot{M}$ = 2$\times$10$^{-9}$ $M_{\odot}$ yr$^{-1}$
for the O9.5~V star $\mu$ Col lies on the green line for weak winds in
Figure~\ref{fig:mdot}.  Note that similar considerations of UV
diagnostics were noted earlier as well \citep{cohen97,cohen08,drew94}.
% Since mass-loss rates for late O-type dwarfs are already quite low
% (and uncertain) compared to post-MS mass loss, the fact that this
% has not yet been incorporated into most stellar evolution
% calculations may not undermine the validity of significantly alter
% those models (single star models have more serious problems to
% contend with anyway).

\subsection{Wolf-Rayet Winds}

The strong winds of WR stars yield spectacular spectra with extremely
strong and broad emission lines, as well as strong excess in the IR
and radio from free-free emission.  \citet{crowther07} has recently
reviewed the properties of WR stars, and typical mass-loss rates for
WN and WC stars are plotted in Figure~\ref{fig:mdot}.  Much of the
discussion of clumping and mass-loss rates of WR stars echoes that of
O-type stars, except for the fact that the effects of clumping were
already known more than a decade ago based on the relative strengths
of electron-scattering wings and emission-line cores, as noted earlier
\citep{hillier91}.  The mass-loss rates of WR stars, like O-type
stars, are line-driven winds and are $Z$ dependent. \citet{vdk05} find
that Fe dominates the driving in WN stars and that they have a
metallicity dependence similar to O-type stars, whereas WC stars have
a somewhat shallower dependence on $Z$ because intermediate-mass
elements from self-enrichment contribute more to the driving as $Z$
drops.
%This may be very important at low $Z$.  It also means that even if
%binary evolution strips a star's H envelope, subsequent evolution
%through the WR sequence (and SN~IIb to Ib to Ic sequence) will still
%be $Z$-dependent.

\subsection{Implications of lower O-star mass-loss rates}

It would appear that the net result of decades of detailed study of
line-driven winds is that, at least concerning stellar evolution, they
don't matter as much as was previously believed (except perhaps for
the most massive stars where proximity to the Eddington limit enhances
the winds; see above).  It is still commonly stated in the literature
that a massive star of $M_{ZAMS} \, = \, 60 \, M_{\odot}$ will shed
half its mass on the main sequence, but if clumping requires us to
reduce mass-loss rates by a factor of 3, then such statements are no
longer true.  For example, a 60 $M_{\odot}$ star will begin the main
sequence with log($L/L_{\odot}$)=5.7 and log($T_{eff}$)=4.7 according
to standard evolutionary models, and $\dot{M}$=10$^{-5.86}$ according
to the \citet{dj88} prescription.  The mass-loss rate will climb
throughout the MS because the luminosity goes up, and so the average
$\dot{M}$ is about a factor of 2 higher.  However, with $\dot{M}$
reduced by a factor of 3 as a standard clumping correction, a 60
$M_{\odot}$ star would ultimately lose only a few $M_{\odot}$ during
the entire 3.5 Myr MS evolution.  This raises the important question
of where WR stars come from in light of lower mass-loss rates.  No
H-free WR stars are known with masses above roughly 20-25 $M_{\odot}$
(see \citealt{crowther07,sc08}), so the lion's share of mass loss is
yet to come.  Stars with $M_{ZAMS} \, = \, 60 M_{\odot}$ don't become
RSGs, and the observed $\dot{M}$ values of steady winds of post-MS
stars at these luminosities (i.e. blue supergiants and quiescent LBVs;
see Figure~\ref{fig:mdot}) are not high enough when combined with the
short duration of these post-MS phases envisioned in single-star
evolution models.  There are only a few options \citep{so06}: (1)
eruptive LBV mass loss makes up the difference, or (2) single stars
with $M_{ZAMS} = 60 M_{\odot}$ don't fully shed their H envelopes
before core-collapse, and binaries are instead responsible for most of
the observed WR stars that may come from these initial masses.  For
stars of lower initial mass $M_{ZAMS}$ = 40-60 $M_{\odot}$, the
problem is worse.  Below $M_{ZAMS}$ $\simeq$ 35 $M_{\odot}$, single
stars should go through a RSG phase and this may help shed the H
envelope to make WR stars.  RSG mass-loss rates are highly uncertain
as well, however (see below).

At higher initial masses above 80-100 $M_{\odot}$, stars pass through
a WNH phase with very strong winds enhanced by high $\Gamma$ values
(see above).  Although the mass loss here may be strong enough to
evaporate the H envelope, stars of such high initial mass would yield
He cores that are more massive than any observed H-poor WR stars.
Since the most massive stars are rare, it is uncertain if this is a
show stopper.  As discussed later, however, it seems easy for known
binaries to account for WR stars, and indeed binary evolution (mass
accretion, mergers) may factor prominently in producing some N-rich WR
stars that may still be H burning.

Even with $\dot{M}$ values reduced by a factor of 3, however,
line-driven winds operating over the entire MS lifetime of O-type
stars may still be quite important in angular momentum loss and the
rotational evolution of massive stars, especially with the possible
aid of magnetic fields.  This important aspect of rotational evolution
is discussed by \citet{langer12}.

\subsection{Metallicity Dependence and Implications for Feedback}

While the winds of OB-type main-sequence (MS) stars may be less
important for a star's evolution than previously thought, having
relatively precise (better than factor of $\sim$2) estimates of
$\dot{M}$ is still desirable to assess the role of wind feedback in
clustered star-forming regions, starbursts, and disk galaxy evolution.
This is because massive stars spend most of their lifetimes as
H-burning O-type stars, and the youngest ages are when the natal ISM
of the star-forming environment is still close to the star and
susceptible to the direct impact of radiation pressure and winds.
Although binary RLOF, eruptive LBV mass loss, and RSG winds remove
more mass from a typical O-type star than line-driven winds, this mass
loss is generally slow and/or cold, it usually happens on a very short
timescale, and it usually occurs late in a star's life, so that the
energy and momentum injection into the surrounding ISM integrated over
the lifetime of the star is far less.  The eventual SNe tend to
explode in a large cavity, and may be less influential as well.  Thus,
for assessing local mechanical feedback from massive stars,
line-driven winds are still an important consideration.

The good news is that, modulo the uncertainty in $\dot{M}$ caused by
clumping, the simple $Z$-dependent scaling of line-driven winds (e.g.,
\citealt{vink01,mokiem07}) is probably reliable enough to estimate the
global contribution of feedback of MS O-type stars from solar to
mildly sub-solar metallicities.  At a given temperature and
luminosity, one expects mass loss to scale as $\dot{M} \propto Z^m$.
Theoretically, \citet{vink01} predict $m$=0.69$\pm$0.10 for O-type
stars, whereas observations suggest $m$=0.83$\pm$0.16
\citep{mokiem07}.  These are in reasonable agreement, although note
that both are steeper than the $Z^{0.5}$ scaling given by KP00 and
adopted by \citet{heger03} (Figure~\ref{fig:heger}) due to the $Z$
dependence of wind speed \citep{vink01}.  These relations have only
been tested to roughly 1/5 $Z_{\odot}$, so extrapolating to
hyper-metal-poor environemnts is still uncertain.

It should be noted that lower mass-loss rates are also important for
assessing the collective {\it radiative} feedback from a cluster or
starburst, because lower values of $\dot{M}$ allow more UV radiation
to escape the wind.  Moreover, the lower mass-loss rates that result
from clumping may indirectly but substantially influence the global
feedback of SNe from a stellar population, since weaker winds during
H-burning could modify the burning lifetime, core size, and end fate
of the star, and hence, the characteristic of the resulting SN
explosion.

\section{DENSE WINDS FROM COOL SUPERGIANTS}

Winds of cool supergiants have received less attention from the
massive-star community than hot-star winds, even though RSG winds are
much stronger (Figure~\ref{fig:mdot}) and far more important for the
evolution of 8$-$35 $M_{\odot}$ stars (i.e., the vast majority of SN
progenitors).  \citet{willson00} has reviewed mass loss from cool
stars, although that paper focussed on lower-masses ($M_{ZAMS} = 1 - 9
\, M_{\odot}$).  The basic physical picture of the mechanism by which
more massive RSG stars lose mass is similar; pulsations lift gas to a
few $R_*$, where the equilibrium temperature becomes low enough
(1000-1500 K) for substantial dust condensation to occur.  Radiation
pressure on newly formed dust (coupled to the gas by collisions in
these dense winds) then takes over and pushes the wind to escape the
star's gravity.  \citet{willson00} also reviewed observational
diagnostics of mass-loss rates in these cool stars.  For lower
mass-loss rates in cool supergiants with coronal winds, UV spectra and
radio emission can be used to investigate the mass-loss rate and other
wind properties (see also \citealt{bennett2010}).  For stronger winds,
the primary methods used to measure the mass loss are with thermal-IR
excess from hot dust and molecular emission.  For extreme mass-loss
rates, one can also use masers \citep{habing96}, and spatially
resolved circumstellar material using a variety of techniques like IR
interferometry \citep{monnier04}.
%or spectroscopy of low-ionization emission lines like Na I and K I
%\citep{bl76,pl94,pl02,smith04}.

The most common prescriptions for RSG mass loss in modern stellar
evolution codes are from the same sources as the old rates for hot
stars \citep{dj88,ndj90}, and they come with comparably large
uncertainty.  This uncertainty must be kept in mind when evaluating
the predictions of single-star evolution models that pass through the
RSG phase.  Some models adopt the significantly higher empirical RSG
mass-loss prescription of \citet{vanloon05}, which is intended for
dust-enshrouded RSGs.
%\footnote{The van Loon rates were derived for LMC metallicity, by
%  scaling the gas:dust masss ratio up by a factor of 2.5 from solar
%  metallicity.}  
Both these relations are shown in Figure~\ref{fig:mdot}.  Depending on
which RSG mass-loss recipe is chosen for a model, RSGs can stay in the
red (lower $\dot{M}$) or be driven to hotter temperatures on the HR
Diagram.  This general behavior has been known for a long time,
discussed in detail with regard to models for the blue progenitor of
SN~1987A \citep{arnett89}.  However, after \citet{smartt09} discussed
that the most massive RSGs apparently do not explode as Type II-P SNe,
new evolutionary tracks that included blueward evolution for RSGs with
initial masses above $\sim$20 $M_{\odot}$ became more common
\citep{ekstrom12}.  Thus, the model outcome of the RSG phase is
sensitively dependent on an uncertain input mass-loss prescription.
Unfortunately, there is, as yet, no well-established quantitative
theoretical prediction for the mass-loss rates of RSG winds or the
detailed physics driving them (including pulsations), as there is for
hot stars.  RSG mass loss is time dependent, and there appears to be a
wide dispersion even for a given luminosity and temperature
\citep{willson00}.  Moreover, the mass loss may not obey any single
prescription throughout the whole RSG evolution of an individual star,
and in fact there is suggestive evidence of this.

\begin{figure}%3	% Figure using psfig.sty
%\centerline{\psfig{figure=hrd_masers.eps,height=20pc}}
\centerline{\psfig{figure=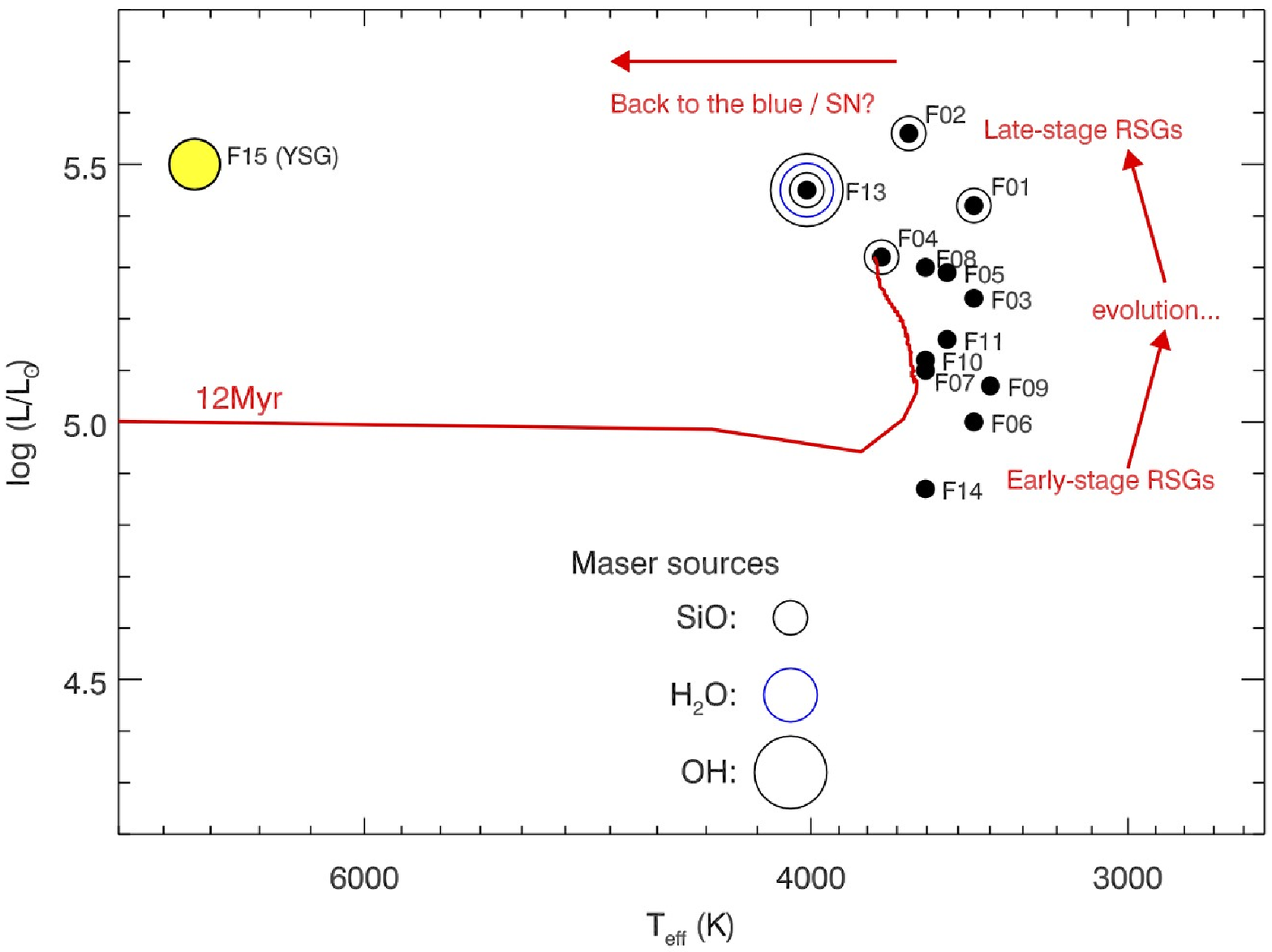,height=20pc}}
\caption{An HR diagram of the cluster RSGC1, with RSGs and a YSG.
  Sources that are circled include maser emission (as noted) from
  dense envelopes, indicating especially strong mass loss.  Adapted
  from \citet{davies08} with permission.}
\label{fig:davies}
\end{figure}

%---------

Observations of massive clusters with numerous RSGs have shed some
important light on this topic.  These provide excellent probes of
massive-star evolution in general --- and RSGs in particular ---
because massive clusters sample a relatively coeval group of stars,
whereas different clusters sample different ages and initial masses of
stars that have reached the RSG phase (see \citealt{davies12} and
references therein).  An example from the cluster RSGC1 is shown in
Figure~\ref{fig:davies} \citep{davies08}.  This compares the locations
of several RSGs in the HR Diagram to a 12 Myr isochrone \citep{mm00},
indicating that these RSGs probably all have initial masses close to
18 $M_{\odot}$.  It is interesting that the RSGs with the strongest
mass loss indicators (traced by H$_2$O, SiO, and OH maser emission;
encircled in Figure~\ref{fig:davies}) are found only at the top of the
RSG branch.  This seems to indicate that the highest mass-loss rates
that lead to self obscuration and maser emission (the sources for
which the van Loon et al. mass-loss rates are appropriate) may be
concentrated toward the very end of the RSG phase when turbulence and
pulsations are most vigorous.  Moreover, this cluster also includes a
yellow hypergiant (YHG) at a luminosity comparable to the most
luminous RSGs with masers.  This provides compelling evidence that
around 18 $M_{\odot}$ or above, enhanced RSG mass loss can indeed
drive these stars toward warmer temperatures at the ends of their
lives.  This, in turn, has implications for connections between
progenitor stars and the types of SNe that they make, as well as their
circumstellar environments into which the SNe explode.

Some of the most luminous RSGs tend to have extremely high mass-loss
rates that cause self-obscuration by dust and strong maser emission.
The best studied Galactic object in this category is VY Canis Majoris,
which has an extended dust-scattering nebula seen in {\it HST} images
\citep{smith01}.  In the case of VY CMa, the density of its nebula
drops off sharply at $\sim$8000 AU from the star, indicating that its
high average mass-loss rate of $\sim$10$^{-3}$ $M_{\odot}$ yr$^{-1}$
has been limited to the previous 10$^3$ yr or so
\citep{shr09,smith01,decin06}.  The nebula around VY CMa is quite
similar to that around the yellow hypergiant IRC+10420, which is
regarded as the prototypical post-RSG because of its maser shell
surviving around such a warm star, and its observed fast blueward
evolution \citep{humphreys97}.  Cases like this provide additional
evidence that strong mass-loss among more luminous RSGs will drive
them on blueward evolutionary tracks.  Some models predict that this
very strong mass-loss phase is short-lived, and enhanced before core
collapse caused by increasingly violent pulsations during C burning
due to a high $L$/$M$ ratio \citep{heger97,yc10,am11}.

\section{SUPER-EDDINGTON WINDS, ERUPTIVE MASS LOSS, AND TRANSIENTS}

Stellar evolution models demonstrate the critical impact of mass loss
on stellar evolution, and we see results of mass loss in the existence
of WR stars and the diversity of SN types.  The uncertainty gets large
as we move toward post-MS phases, as we have seen in the case of RSG
mass-loss rates, and the probably dominant role of binaries (next
section).  The uncertainty is worst at the highest stellar
luminosities, where mass-loss is strongest, and where systems observed
in detail are few.  Proximity to the Eddington limit can enhance the
strength of a steady wind, or worse, it can make the star unstable,
potentially leading to violent eruptive or explosive mass loss
associated with transient events that can dramatically change the star
in a short time.  For extreme cases in very massive stars, a single
eruptive event can remove more mass in a few years than is shed during
its MS evolution.  It is therefore sobering to recognize that {\it
  none} of these effects are included in current generations of
stellar evolution models.

In section 2.3 we briefly mentioned the role of the Eddington ratio,
$\Gamma$, in enhancing the relatively steady mass-loss rates of very
luminous H-burning stars, like the WNH stars.  Here we focus on the
more extreme cases where the high $\Gamma$ leads to instability in the
star and highly time-dependent mass loss, or where advanced nuclear
burning stages may play a role \citep{am11}.  When this variability is
observed, the stars are designated as LBVs, or LBV candidates if they
are suspected to be dormant versions of the same stars.

\subsection{LBVs: History and Phenomenology}

The most dramatic instability arising in post-MS evolution is the
class of objects known as luminous blue variables (LBVs).  These were
recognized early as the brightest blue irregular variables in nearby
galaxies \citep{hs53,ts68}, and these classic examples were referred
to as the ``Hubble-Sandage variables''.  Famous Galactic objects like
P Cygni and $\eta$ Carinae had spectacular outbursts in the 17th and
19th centuries, respectively, but their connection to other eruptive
massive stars was unclear.  \citet{conti84} recognized that many
different classes of hot, irregular variable stars in the Milky Way
and Magellanic Clouds were probably related to the Hubble-Sandage
variables and to $\eta$ Car and P Cygni, so he grouped them together
as ``LBVs''. The LBVs are a rather diverse class, consisting of a wide
range of irregular variable phenomena
\citep{hd94,vg01,svdk04,smith11a,vdm12,clark05}.  Their initial masses
are uncertain, but comparing their luminosities to single-star
evolution tracks suggests initial masses greater than 25 $M_{\odot}$
\citep{svdk04}.  LBVs are defined by their irregular eruptive
variability.  There are, however, stars that spectroscopically
resemble LBVs in their quiescent state, but which have not (yet) been
observed to show the signature variability of LBVs; these are often
called LBV candidates, and they are usually of spectral type Ofpe/WN9
or early B supergiants.  It is not known if LBVs pass through long
dormant periods, and if they do, the duty cycle is unknown.  As noted
below, the detection of a dense circumstellar shell is often taken to
indicate a prior giant outburst.

{\bf S Doradus phases.} S~Dor outbursts are seen as a visual
brightening that occurs when the peak of the star's energy
distribution shifts from the UV to visual wavelengths.  The increase
in visual brightness (i.e. 1--2 mag, typically) corresponds roughly to
the bolometric correction, so that hotter stars exhibit larger
amplitudes.  In their quiescent states, LBVs have apparent
temperatures that increase with increasing luminosity: they often
appear as Ofpe/WN9 stars at high luminosity, or early/mid B
supergiants at the lower luminosity end \citep{svdk04,hd94}.  Visual
maximum occurs at a constant temperature of $\sim$8000 K, causing the
star to resemble a late F supergiant.  S Dor events were originally
proposed to occur at constant bolometric luminosity \citep{hd94}, but
quantitative studies do reveal varitions in $L_{Bol}$ \citep{groh09}.
The traditional explanation for the apparent temperature change was
that the star dramatically increases its mass-loss rate, driving the
wind to very high optical depth and causing a pseudo photosphere
\citep{davidson87,hd94}.  However, quantitative spectroscopy revealed
that the measured mass-loss rates in outburst do not increase enough
to cause a pseudo photosphere \citep{dekoter96}, and that the
increasing photospheric radius is therefore more akin to a pulsation.
A possible cause of this inflation of the star's outer layers may be
near-Eddington luminosities in the sub-surface Fe opacity bump
\citep{grafener12,guzik12}.  S Dor eruptions of LBVs are therefore
{\it not major mass-loss events}.  However, the average mass-loss rate
in a wind throughout the LBV phase (quiescent or not) is about an
order of magnitude higher than for O-type stars of comparable
luminosity (Figure~\ref{fig:mdot}).

{\bf LBV Giant eruptions.}  The most pronounced variability attributed
to LBVs is their so-called ``giant eruptions'', in which stars are
observed to increase their bolometric luminosity for months to years,
accompanied by extreme mass loss \citep{hds99}.  The best studied
example is the Galactic object $\eta$ Carinae, providing us with its
historically observed light curve \citep{sf11}, as well as its complex
ejecta that contain 10-20 $M_{\odot}$ and $\sim$10$^{50}$ ergs of
kinetic energy \citep{smith06,smith03}.  Light echoes from the Great
Eruption of $\eta$ Carinae have just recently been discovered
\citep{rest12}, and their continued study with spectroscopy may modify
long-held ideas about LBVs.  A less well-documented case is P Cygni's
1600 AD eruption, for which a much smaller ejecta mass of 0.1
$M_{\odot}$ has been measured \citep{sh06}.  P Cyg's nebula has an
expansion speed of $\sim$140 km s$^{-1}$ \citep{sh06,barlow94}, with
an implied total kinetic energy of a few 10$^{46}$ ergs.  P Cyg and
$\eta$ Car are the only two cases of observed LBV giant eruptions
where the ejected mass has actually been measured, because they have
spatially resolved shell nebulae ejected in the events.  With
decade-long durations, the implied mass-loss rates are at least 0.01
$M_{\odot}$ yr$^{-1}$ and 1 $M_{\odot}$ yr$^{-1}$ for P Cyg and $\eta$
Car, respectively.  These rates are too high to be driven by
traditional stellar winds because the material is opaque
\citep{so06,owocki04}.  Dust formation in LBV eruptions also points
toward eruptive mass loss \citep{kochanek11b}.

{\bf Extragalactic SN Impostors.}  LBV giant eruptions are rare, so
our only other observed examples are a few dozen found in nearby
galaxies \citep{vdm12,smith11a}.  Due to their serendipitous discovery
in SN searches, they are sometimes called ``SN impostors''.  Other
names include ``Type V'' SNe, ``$\eta$ Car analogs'', and various
permutations of ``intermediate luminosity transients''.  These have
peak absolute magnitudes of $-$11 to $-$15 mag
\citep{vdm12,smith11a}. Typical expansion speeds observed in outburst
spectra are 100-1000 km s$^{-1}$ \citep{smith11a}, although lower
speeds can be seen along the line of sight if the ejection speed is
latitude dependent \citep{smith06}.  A realization in the past decade
is that there is wide diversity among the SN impostors and their
progenitors; some events that resemble LBV eruptions may actually
arise in lower-mass progenitor stars \citep{thompson09,prieto08}.  A
review of the lower-mass analogs of LBV giant eruptions is beyond the
scope of this article, but the fact that lower-mass stars may
experience similar transient events casts doubt on the long-held
belief that these eruptions result from high luminosities near the
Eddington limit (see below).

{\bf LBV winds.}  Most LBVs exhibit strong emission lines in their
visual-wavelength spectra, similar to WR stars but with narrower
widths and stronger H lines.  Wind speeds are typically 100-600 km
s$^{-1}$, reflecting the lower escape speed of BSG stars as compared
to 1000-2000 km s$^{-1}$ in more compact O and WR stars. The wind
mass-loss rates implied by quantitative models of the spectra
typically range from 10$^{-5}$ to 10$^{-4}$ $M_{\odot}$ yr$^{-1}$
\citep{groh09,svdk04,vdk02,dekoter96}, or even 10$^{-3}$ $M_{\odot}$
yr$^{-1}$ in the extreme case of $\eta$ Car \citep{hillier01}.  These
LBV wind mass-loss rates are indicated in Figure~\ref{fig:mdot}.  LBV
winds are strong enough to play an important role in the evolution of
the star if the LBV phase lasts more than 10$^5$ yr, and eruptions
further enhance the mass loss.  Other stars that exhibit similar
spectra but are not necessarily LBVs include WNH stars, Ofpe/WN9
stars, and B[e] supergiants, which overlap on the HR Diagram.

\begin{figure}%3	% Figure using psfig.sty
%\centerline{\psfig{figure=nebulaLBV.eps,height=20pc}}
\centerline{\psfig{figure=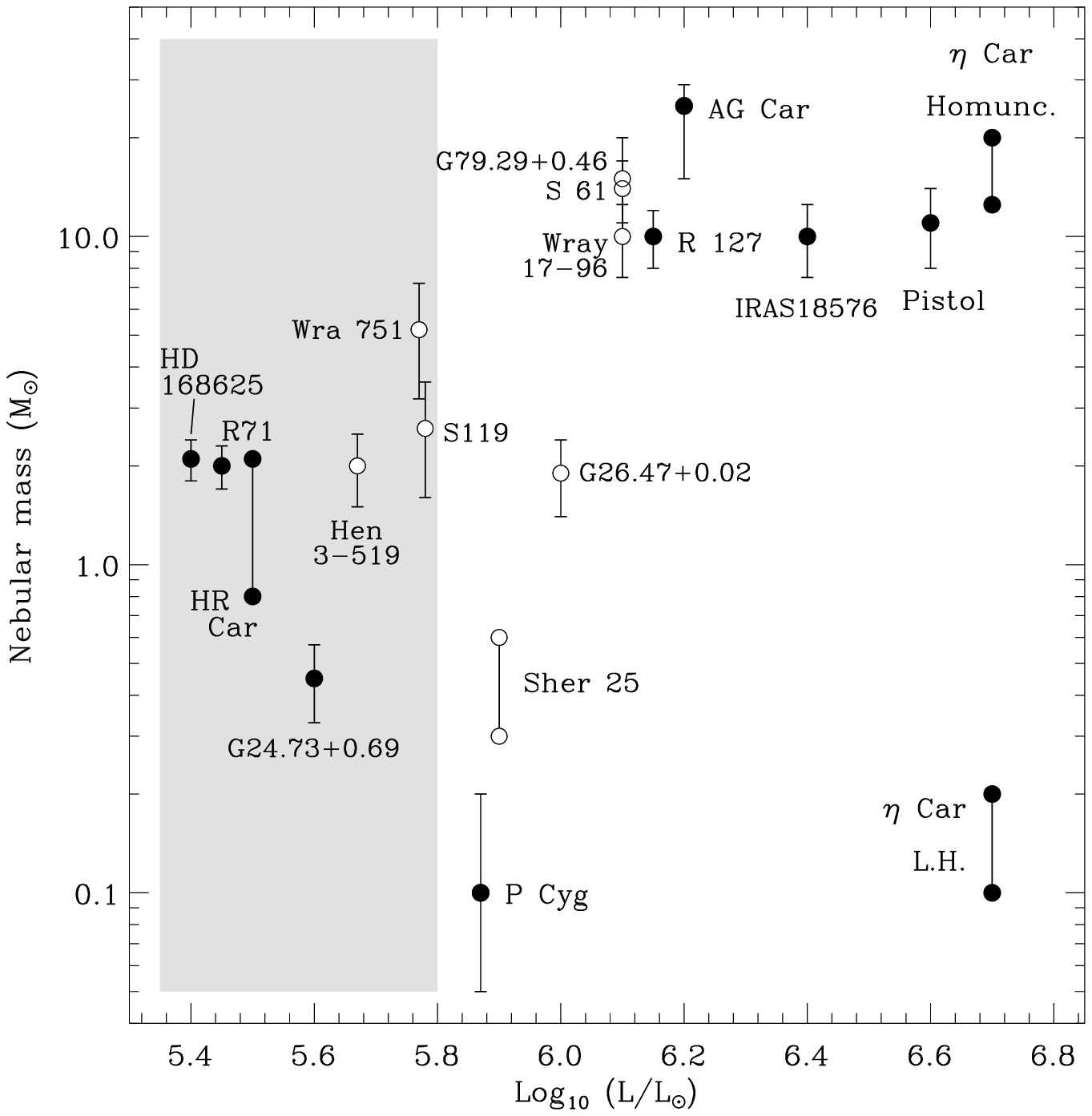,height=20pc}}
\caption{Masses of circumstellar shells around LBVs and LBV-like
  stars, as a function of luminosity, from \citet{so06}.  The left
  side of the plot (grey box) corresponds to stars below
  log($L/L_{\odot}$)=5.8, so these LBVs could be post-RSGs and the
  nebular mass could have been ejected in the RSG phase and swept up
  into a shell.  Objects on the right must have ejected their massive
  shells in giant LBV eruptions.  There may be many lower-mass shells
  that are hard to detect around the very bright central stars.}
\label{fig:nebulae}
\end{figure}

{\bf Circumstellar shells.}  Many LBVs have spatially resolved
circumstellar shells that are fossils of previous eruptions.  Stars
that resemble LBVs spectroscopically and have massive shells, but have
not been observed to exhibit LBV variability, are called LBV
candidates, as noted earlier.  LBV circumstellar shells are extremely
important, as they provide the only reliable way to estimate the
amount of mass ejected in an LBV giant eruption.  A large number of
LBVs and candidates in the Milky Way and Magellanic Clouds are
surrounded by massive shell nebulae
\citep{wachter10,gvaramadze10,clark05,so06}.  Thus, eruptive LBV mass
loss is inferred to be important in the late evolution of massive
stars.  Masses of LBV nebulae occupy a very large range from $\sim$20
$M_{\odot}$ at the upper end down to 0.1 $M_{\odot}$, although even
smaller masses become difficult to detect around bright central
stars. In some cases a very large range in mass is seen in multiple
shells around the same star, as for $\eta$ Car \citep{smith05}, so
there is no clear one-to-one correlation of shell mass and stellar
luminosity, although there may be such a relation for the most massive
shell a star can eject.  Masses for a collection of LBV shells are
shown in Figure~\ref{fig:nebulae}, compiled by \citet{so06}.  To
compare with steady winds, LBV giant eruption mass-loss ``rates'' in
Figure~\ref{fig:mdot} are shown by dividing LBV shell masses by the
$\sim$10 yr duration in the eruptions of $\eta$~Car and P~Cyg,
although the true instantaneous $\dot{M}$ may be higher.  Dynamical
ages of the shells around LBVs/candidates range from 10$^2$ yr to
several thousand years.  It is, however, difficult to use this age as
an indication of the duration of the LBV phase, since expanding shells
can decelerate.  The duration of the LBV phase may be further
compliated, since it may vary with initial mass and may depend on
details of binary evolution.

\subsection{Physics of Eruptive Mass Loss}

Although the violent variability of LBVs has been known for a century
or more, the search for a physical theory of LBV eruptions is still in
early stages.  Most work so far has concentrated on how to lift
material off the star, or on tighter observational constraints on the
mass, speed, and energy of outbursts.  In terms of driving the mass
loss, two broad classes of models have developed: super-Eddington
winds and explosions.  Both may operate at some level, but neither of
these addresses the deeper question of what initiates LBV eruptions in
the first place.  Ideas for the underlying trigger are still
speculative.

\subsubsection{Super-Eddington Winds}

Traditionally, LBV giant eruptions have been discussed as
super-Eddington (SE) winds driven by a sudden unexplained increase in
the star's bolometric luminosity
\citep{hd94,hds99,shaviv00,owocki04,so06}.  This is motivated mostly
by the fact that $\eta$ Car's Great Eruption had an observed
luminosity that indicated $\Gamma\simeq5$ for about a decade or more,
and that extragalactic SN impostors show similar high luminosities.
With $\Gamma$ values substantially above unity, one naturally expects
strong mass loss, but the detailed physical picture of such winds is
not obvious.  SE winds are expected to be very dense but porous,
allowing the star's atmosphere to remain in steady state while
exceeding the classical Eddington limit \citep{shaviv00,owocki04}.  An
important point is that the mass loss is so strong that the high
density in the wind causes UV absorption lines to be saturated.
Therefore, SE winds are driven by photon momentum tranferred to gas
through electron scattering opacity and not line opacity
\citep{owocki04}.  This makes SE wind mass loss essentially
independent of metallicity, which may allow this mode of mass loss to
operate in Pop III stars \citep{so06}.  

Numerical simulations of continuum-driven SE winds show complex
structure with both infall and outflow \citep{vanmarle08,vanmarle09},
confirming expectations that these winds should be highly
inhomogeneous \citep{shaviv00,owocki04}.  SE winds may also account
for the bipolar shape of nebulae around LBVs like $\eta$ Car if the
star is a rapid rotator, since equatorial gravity darkening will lead
to a higher $\dot{M}$ and faster speed in the polar wind
\citep{owocki96,do02}.

Open questions surrounding SE winds are whether the mechanism can
supply the mass loss in the most extreme observed cases, and what
initiates the SE phase.  Current estimates of the mass lost in $\eta$
Car's 19th century eruption are of order 15 $M_{\odot}$ or more
\citep{smith03,sf07}.  SE winds can in principle cause that much mass
loss averaged over 20 years \citep{owocki04}, but those same models
predict relatively slow outflow speeds.  This makes it hard to explain
the nebula around $\eta$ Car, with most of the mass moving at speeds
of 500-600 km s$^{-1}$.  Moreover, $\eta$ Car also shows a smaller
mass of extremely fast material moving at 5000 km s$^{-1}$ or more
\citep{smith08}, which is hard for a SE wind model to achive while
driving such a large amount of mass.  SE winds are still viable for
most LBVs, which are less extreme than $\eta$ Car's 19th century
eruption.  As for the extra radiative energy output that initiates the
SE wind, this is not known.  Lastly, SE winds are assumed to be
launched from the surface of the star, but it is also not yet clear if
the star's interior can remain stable at $\Gamma=5$ for more than a
decade.

\subsubsection{Explosions}

There is growing observational evidence that some giant LBV eruptions
may be non-terminal hydrodynamic explosions.  Part of the motivation
for this is based on detailed study of $\eta$ Carinae, which has shown
several signs that the 1840s eruption had a shock-powered component to
it.  This includes estimates of the ratio of total ejecta kinetic
energy to the integrated radiated energy of $E_k/E_{rad}>3$, which is
hard for a radiation-driven wind to achieve (although perhaps not
impossible with extreme photon-tiring; \citealt{owocki04}).  The very
thin walls of the nebula indicate a small range of expansion speed
\citep{smith06}, which is easiest to achieve from compression in a
shock.  Lastly, observations show extremely high-speed ejecta moving
at 5000 km s$^{-1}$, which seems impossible to achieve without a
strong blast wave \citep{smith08}. A number of extragalactic LBV-like
eruptions show spectra that closely resemble shock-powered Type~IIn
SNe and also show evidence for extremely fast ejecta that may signify
a shock-powered event, such as the precursor outbursts of SN2009ip
\citep{smith10,foley11}.

One normally expects sudden, hydrodynamic events to be brief (i.e., a
dynamical time), which at first may seem incompatible with the
decade-long Great Eruption of $\eta$ Car.  However, an explosion
followed by CSM interaction can generate a high sustained luminosity,
as in core-collapse SNe~IIn.  \citet{smith13a} showed that a
shock-powered event with CSM interaction could account for the
1845-1860 light curve of $\eta$ Car, using a SN~IIn-type model but
with lower explosion energy.  The resulting slower shock speed from a
sub-energetic explosion (10$^{50}$ instead of 10$^{51}$ ergs) produces
a lower CSM interaction luminosity compared to a core-collapse SN~IIn,
and takes much longer to expand through the CSM.  The duration of the
event is determined by the outer extent of the dense CSM --- in
principle, a CSM-interaction powered LBV eruption might continue for
several decades, or it could last only 100 days \citep{fa77},
depending on the extent of the dense pre-explosion wind.  Since
shock/CSM interaction is such an efficient way to convert explosion
kinetic energy into luminosity, it is plausible that many of the SN
impostors with narrow emission lines may be powered in this way.  The
shock model would help explain the wide observed diversity of SN
impostors \citep{smith11a,vdm12}.  The catch is that even this model
requires {\it something else} to create the dense CSM into which the
shock expands, which may be where super-Eddington winds or binary
interaction play an important role.
%The physical benefit of this model is that the demands on the
%super-Eddington wind are relaxed to a point that is physically
%achievable; instead of driving 10 $M_{\odot}$ in a few years (as for
%$\eta$ Car), the SE wind can provide roughly half the mass spread over
%several decades or a century.  Also, the wind can be slow as we might
%expect for SE winds \citep{owocki04}, whereas the kinetic energy comes
%from the explosion.

\subsubsection{Eruption Triggers}

The reason for the onset of an LBV eruption and its power source
remain unanswered for either mechanism.  In the SE wind model, even if
the wind can be driven at the rates required, we have no underlying
physical explanation for why the star's bolometric luminosity suddenly
increases by factors of 5-10, and we don't know how the star's
envelope would process that high energy flux.  In the explosion model,
the underlying trigger for an explosive event is unknown.  In either
case, {\it something} must inject a large amount of extra energy
(10$^{48}$ -- 10$^{50}$ ergs) into the star's interior in a highly
time-dependent way.  There have been a number of physical mechanisms
discussed in connection with LBVs, recently reviewed in detail by
\citet{smith11a}.  In brief, there is no clearly favored explanation,
but some ideas appear to be ruled out based on the required energy and
mass budgets. These can be thought of in two broad categories:
instability and energy deposition.

{\it Envelope Instability.}  LBVs are massive stars that are in close
proximity to the Eddington limit.  Consequently, their loosely bound
envelopes may be susceptible to strange-mode instabilities
\citep{gk93,glatzel99}, runaway mass loss (a.k.a. the ``Geyser''
model; \citealt{maeder92,hd94}), or the critical rotation limit
(a.k.a. the ``$\Omega$ limit''; \citealt{langer98,langer12}).  While
these may help explain some of the irregular variability seen in S Dor
outbursts of LBVs, they are unsatisfactory explanations for giant LBV
eruptions.  This is because the total mass ejected can be much more
than the small mass in the outer H envelope where the relevant
instabilities reside, and LBV eruptions can have substantially more
kinetic energy than the total thermal energy in the star's envelope.

{\it Energy deposition.}  A large amount of extra energy can be
deposited deep in a massive star's envelope by a number of suggested
mechanisms, including unsteady burning \citep{sa13}, the pulsational
pair instability \citep{whw02,woosley07}, other explosive shell
burning instabilities \citep{dessart10,sa13}, wave-driven mass loss
\citep{qs12,sq13,ma07}, and stellar collisions or mergers in a binary
system \citep{smith11,sa13,podsiadlowski10}.\footnote{Soker and
  collaborators have discussed a model to power the luminosity in LBV
  eruptions using accretion onto a companion star \citep{ks09}, but
  this invokes an eruption to provide the mass that is then accreted;
  it does not explain what initiates the mass loss from the primary in
  the first place.}  While any of these provides a plausible result,
the main criticism for explaining LBV eruptions is that the mechanisms
which are related to late nuclear burning instabilities are expected
to occur only in the few years preceeding core collapse.  However,
many LBVs with massive shells appear to have survived for
10$^2$-10$^4$ yr after a giant eruption.  Such an objection turns into
an advantage, however, in the case of violent pre-SN eruptions needed
for SNe~IIn (see below).

Research on LBV eruptions and pre-SN eruptions is actively ongoing,
and it is a major unsolved problem in astrophysics.  Observations
demonstrate that these events do occur, and the mass budget involved
can dominate or significantly contribute to the total mass lost by a
massive star.

\subsection{The Role of LBV Mass Loss in Stellar Evolution}

If this section were to review the influence of LBV eruptions when
they are included in stellar evolution models, it would be a very
short section.  Without exception, no current stellar evolution models
account for LBV giant eruptions. This is because we don't know how to
include them properly.  Observationally, we don't have reliable
estimates of the typical mass ejected as a function of the star's
initial mass, how many repeating eruptions occur for a given star, or
the duty cycle of eruptions.  Theoretically, we don't know the
underlying physical mechanism(s), when they occur during the evolution
of a star, or how they should vary with $M_{ZAMS}$.

The traditional view of LBVs, which emerged in the 1980s and 1990s, is
that they correspond to a very brief transitional phase of evolution,
when the massive star moves from core H burning to core He burning
\citep{hd94}.  A typical monotonic evolutionary scheme is:

\smallskip 

\noindent 100 $M_{\odot}$: O star $\rightarrow$ Of/WNH $\rightarrow$
LBV $\rightarrow$ WN $\rightarrow$ WC $\rightarrow$ SN Ibc

\smallskip

\noindent In this scenario, the strong mass-loss experienced by LBVs
is important for removing what is left of the star's H envelope after
the MS, leaving a WR star following the LBV phase.  The motivation for
a very brief phase comes from the fact that LBVs are extremely rare:
the duration of the LBV phase is thought to be only a few 10$^4$ yr
\citep{hd94}.

However, a number of inconsistencies have arisen with this standard
view.  The very short inferred LBV lifetime depends on the assumption
that the observed LBVs occupy the whole transitional phase.  In fact,
there is a much larger number of blue supergiant stars that are not
seen in eruption --- these are the LBV candidates.  Examining
populations in nearby galaxies, Massey et al. (2007) find that there
are more than {\it an order of magnitude} more spectroscopically
similar LBV candidates than there are LBVs confirmed by their
variability. If LBV candidates are included, then the average LBV
phase rises from a few 10$^4$ yr to several 10$^5$ yr.  This is
comparable to the whole He burning lifetime of very massive stars,
making it impossible for LBVs to be mere {\it transitional} objects.
Massey (2006) has pointed to the case of P Cygni as a salient example:
its 1600 A.D.\ giant LBV eruption was observed and so we call it an
LBV, but it has shown no eruptive LBV-like behavior since then. If the
observational record had started in 1700, then we would have no idea
that P Cygni was an LBV.

Another major issue is that we have growing evidence that LBVs or
something like them (massive H-rich stars with high mass loss, N
enrichment, slow 100-500 km s$^{-1}$ winds, massive shells) are
exploding as core-collapse SNe while still in an LBV-like phase (see
below).  This could not be true if LBVs are only in a brief transition
to the WR phase, which should last another 0.5-1 Myr.

A prolonged LBV phase, and indeed any very massive stars above
$M_{ZAMS}$ = 40 $M_{\odot}$ making it to core collapse with H
envelopes still intact at $Z_{\odot}$, is in direct conflict with the
single-star evolution models.  The evolutionary state and basic nature
of LBVs is therefore still quite uncertain.  This is problematic for
our understanding of massive star evolution, in which mass-loss is a
key ingredient, since LBVs have the highest known mass-loss rates of
any stars (Figure~\ref{fig:mdot}).

\subsection{Low Metallicity}

While we don't yet know the root cause of LBV eruptions, we do know
that the huge observed mass-loss rates demand that the mechanism
imparting momentum to the ejecta is not a $Z$-dependent wind, because
the outflowing material is very optically thick and lines are
saturated.  LBV eruptions must be either continuum-driven SE winds or
hydrodynamic events (violent pulsations, explosions), and both of
these are relatively insensitive to $Z$.  Since this mode of eruptive
mass loss may actually dominate the total mass shed by a massive star
at $Z_{\odot}$, there is not yet any reason to think it won't also
work in the early universe. This may be important for Pop III stars,
since they are argued to have been preferentially very massive.  There
may, of course, be some unrecognized way that metallicity creeps into
the problem (i.e. Fe opacity bumps, some $Z$ dependence of explosive
burning, etc.), but this has not yet been investigated.  LBVs have
been identified in nearby low-$Z$dwarf galaxies \citep{it09,izotov11},
in addition to the very nearby case of HD~5980 in the SMC, and the
LBV-like eruptions that precede SNe~IIn (see below) often occur in
dwarf galaxies. Thus, there is empirical evidence that low metallicity
does not inhibit eruptive mass loss.

\section{BINARY MASS TRANSFER, MERGERS, AND MASS LOSS}

At massive-star conferences, a common refrain heard upon completion of
a talk about single-star evolutionary models is: ``What about
binaries?''  This is often followed by an uncomfortable silence,
shrugging of shoulders, nervous laughter by the speaker, or
intervention by the session chair to move to a serious question.

Actually, this {\it is} a serious problem.  Exclusion of complicated
binary effects and a focus on single-star evolution models is valid,
in principle, because indeed we must start by having a foundation in
understanding single stars.  However, comparing single-star models to
the observed {\it statistical} properties of stars -- and then using
these diagnostics to inform the validity of assumptions in those
single-star models -- can lead to serious errors if binaries make a
significant contribution to observed distributions.  The most critical
influence of binaries on the observed distribution of massive stars
comes through the process of mass loss and mass transfer via RLOF,
which exceeds the influence of stellar winds and rotation for most
(and possibly all) intial masses.

\subsection{Massive Stars are Mostly in Binaries}

Based on work in the past decade, we now have secure evidence that the
binary fraction is not only high among massive stars, but that {\it
  the large majority of massive stars (roughly 2/3-3/4) reside in
  binary systems that have orbital periods short enough that the stars
  will interact and exchange mass (or merge) during their life}
\citep{sana12,se11,mason09}.  There have been several monitoring
campaigns to measure the observed spectroscopic binary fraction among
massive stars in clusters using radial velocities, which typically
find an {\it observed} binary fraction of 20-60\%
\citep{sana12,sana08,sana09,kk12,kiminki12,chini12,mahy09,evans06,debecker06,gm01,gies87}.
However, these observed binary fractions are only a lower limit,
because they must be corrected for the spectroscopic binary systems
that are missed because of low inclination, periods that are too long
compared to the observational cadence or have high eccentricity, or
low-mass campanions that are more difficult to detect.  Some fraction
of these have orbital separation small enough that the stars will
actually interact and exchange mass, typicaly determined by the
maximum radius of a RSG (roughly 5 AU) or LBVs in outburst (a few AU)
depending on the initial masses of the stars \citep{kk12}.

Recent estimates including the results from several young clusters
suggest that, when corrected for obervational bias, the fraction of
massive stars in binary systems whose orbital period is so short that
the stars must exhange mass or merge is something like 3/4
\citep{sana12,kf07,kk12}.  Of the total population of massive stars,
\citet{sana12} estimate that $\sim$1/4 will merge, $\sim$1/3 will have
their H envelopes stripped before death, and $\sim$14\% will be spun
up by accretion, whereas only about 1/4 of massive stars are actually
effectively single (including stars in wide binaries).  This means
that binary RLOF is not just a factor that we should perhaps consider,
but that it must {\it dominate} the observed effects of mass loss and
mixing seen in massive stars.  It is also likely that the vast
majority of the most rapidly rotating stars, including potentially
{\it all} Be stars, result from interaction with a companion star
\citep{demink13} rather than single stars born with a very high
rotation rate.  This, in turn, suggests that influence of rotation and
rotational mixing in current single-star models may be overstimated
\citep{demink13,vanbeveren09}.

\subsection{Physics of Mass Loss in Binaries}

From the ZAMS until the H envelope is substantially stripped or removed
completely, a massive star will tend to have its radius increase due
to interior evolution.  This begins on the MS as stars steadily become
more luminous and slightly cooler, and the radius then expands more
severely in post-MS phases.  In a binary system, significant mass
transfer begins when the primary star expands to a size where its
photosphere crosses the inner Lagrange point (L1).  

The onset of RLOF therefore depends sensitively on the initial orbital
separation; it may occur while the primary is still on the main
sequence (orbital periods of a few days) or when it expands to a much
larger size in post-MS supergiant phases (orbital periods of 10s of
days or longer).  RLOF on the MS is referred to as Case A, whereas
Case B is for RLOF during H shell burning and Case C for core He
burning \citep{langer12,petrovic05,podsiadlowski92}.  Mass transfer
becomes increasingly unstable and rapid from Case A to C, in some
cases involving dynamical events, common envelopes, or inspiral and
mergers.  The mass transfer rate also depends sensitively on the mass
ratio $q=M_2/M_1$ ($q\le1$), with increasing rates for lower values of
$q$. For Case A, the orbit widens and the mass transfer rate drops as
$q$ approaches unity.

Mass-transfer rates (and hence, mass-loss rates from the primary) can
be very high, and can far exceed any mass-loss rate for a line-driven
wind (see Figure~\ref{fig:mdot}).  The detailed physics of mass
transfer in RLOF is quite complicated and time-dependent.  There is
much remaining uncertainty about the proper treatment of contact
systems, as well as their mass and angular momentum loss.  A common
prescription for the strongest phases of RLOF mass transfer is to
assume that the mass transfer is limited by the thermal timescale of a
massive star with a radiative envelope:

\begin{displaymath}
\dot{M} = (M_0 - M_{WR})/ \tau_{KH}
\end{displaymath}

\noindent where $\tau_{KH}$ is the thermal (Kelvin-Helmholtz)
timescale of the envelope, $M_0$ is the initial mass, and $M_{WR}$ is
the mass of the resulting WR star.  Mass-transfer rates can therefore
be the highest for more massive and more luminous stars, which have
short thermal timescales.  Resulting mass transfer rates can be
extremely high; during fast Case A or Case B and C phases, mass-loss
rates can be of order 10$^{-3}$ $M_{\odot}$ yr$^{-1}$ or higher
\citep{langer12,ts00}, although there is a wide range (only order of
magnitude values are represented in Figure~\ref{fig:mdot}).  With
thermal timescales of order 10$^4$ yr, binary RLOF is therefore
capable of quickly removing almost the entire H envelope of a massive
star and leaving behind a WR star \citep{petrovic05}.  When RLOF ends,
the star's radius will shrink and there will be a small residual H
layer on the star.

The physics and observed phenomenology of the RLOF phase and common
envelopes can be very complicated \citep{ts00}, and could probably
fill several review articles.  The main uncertainties are in the
degree to which mass transfer is
conservative\footnote{``Conservative'' mass transfer means that no
  significant mass is lost from the binary system in RLOF.}
\citep{cantiello07,demink07}, the related question of how the mass
gainer responds to the added angular momentum that may lead to
critical rotation \citep{demink13}, and the consequent orbital
evolution.  The most important point for our purpose here is to
concentrate on the net result: large amounts of mass are ``instantly''
(relative to nuclear burning timescales) stripped from one star, and
much or all of this is accreted by the other.  The envelope stripping
of the primary is similarly efficient to the also ``instantaneous''
removal of $\sim$10 $M_{\odot}$ in giant eruptions of LBVs (see
above).  Indeed, their mass-loss rates are suspiciously similar in
Figure~\ref{fig:mdot}, and it remains possible that some LBV giant
eruptions are extreme mass transfer or merger events (many LBVs or LBV
candidates have only a single massive CSM shell).
\citet{podsiadlowski10} have discussed extreme cases where mixing of
fresh fuel into deeper layers during a binary merger might lead to
explosive burning and removal of the H envelope, reminiscent of some
ideas for explosive LBV giant eruptions.

There are few observational constraints on mass loss and mass transfer
rates in binaries undergoing RLOF.  Since the most active phases are
very brief ($\sim$10$^4$ yr; less than 1\% of a massive star's life),
systems undergoing strong RLOF at any given time are rare.  (The
longer-lasting systems in slow Case A RLOF are more common, but have
lower mass-transfer rates; see \citealt{demink07}.)  One well-studied
example of a short-period system caught in the phase of fast Cas A
mass transfer is the 11-day eclipsing binary RY Scuti.  It has
component masses of roughly 8 and 30 $M_{\odot}$, with a mass-gainer
surrounded by an opaque disk \citep{grundstrom07} and a spatially
resolved toroidal nebula \citep{smith11c}.  A noteworthy object for
wider orbital separations (probably Case C) is the famous YHG star
HR~5171A, which \citet{chesneau13} recently resolved as a mass
transfer binary using interferometry.  This system hints that some of
the YHGs may actually be wide binary systems, where RLOF truncates
further redward evolution of the primary, rather than being products
of single-star RSG mass loss.  Since the companion was not discovered
until it was resolved with interferometry, this also demonstrates that
these binaries may be easily hidden.  Interestingly, \citet{prieto08b}
report the discovery of a rare extragalactic YSG eclipsing binary,
which may provide a similar indication.  Aside from these rare cases,
much of our empirical understanding of RLOF therefore comes from
studying the more common post-RLOF binaries (WR+OB systems).  Their
inferred histories depend on a number of assumptions, however
\citep{hellings84}.

\subsection{Binary Evolution Models and Population Synthesis}

The idea that binary RLOF strips a star's H envelope to dominate the
production of WR stars and SNe Ibc is an old one \citep{pac67}.  As
noted in the introduction, this view fell out of favor due to the
estimated strength of radiatively driven winds, but is now
experiencing a resurgence due to lower wind mass-loss rates and very
high observed binary fractions.  As such, it is now clear that the
complexity of RLOF and its many varying parameters are a necessary
evil to consider.  Some advocates for the importance of binaries may
point out that this was true all along (e.g., \citealt{vanbeveren98}).

Before and after binary stars interact, they behave largely as single
stars, and so binary stellar evolution models come with all the
assumptions and uncertainties that go into single-star models, like
the treatment of convection, assumptions about convective overshoot,
the importance of rotational mixing and angular momentum diffusion,
and of course wind mass loss \citep{mm00,whw02}.  Binary evolution
introduces additional parameters \citep{langer12}.  The total mass
lost or transferred in RLOF, its transfer rate, the amount of angular
momentum lost/transferred, and time dependence of these are influenced
by several intial conditions: 1) the primary/secondary initial mass
ratio $q$, 2) the primary star's intial mass, 3) the initial orbital
separation, 4) the orbit eccentricity, and 5) the relative wind
mass-loss rates of both stars.  It may also matter if a system has
additional multiplicity, but this is usually ignored.  Calculating
grids of detailed binary stellar evolution that explore this parameter
space and also provide a detailed treatment of rotation and mixing in
the stars would require a large fraction of the computing power on
Earth.  Therefore, present state-of-the-art binary models and
population synthesis must make simplifying assumptions
\citep{langer12}.
%%, such as not calculating the evolution of a mass gainer until RLOF
%ends, shortcuts for the treatment of angular momentum, assumptions
%about maximum mass-transfer rates to avoid the need to calculate
%hydrodynamics, ignoring the rotation of one or both stars, etc.
Binary population synthesis models have demonstrated that RLOF can
naturally account for many of the observed statistical distributions
of stellar types, as well as the relative rates of various types of
SNe \citep{podsiadlowski92,petrovic05,eldridge08,es09,
  yoon10,dessart11,sana12,vanbeveren98,vanbeveren07}.  Earlier studies
made inferences about what the interacting binary fraction would need
to be in order to account for observations.  If new estimates of the
very high binary fraction are correct, it now seems unavoidable that
binary evolution will dominate the observed populations of WR stars
and SNe~Ibc (sections 6.1 and 6.2).

%%%%%%%%-------------------------

The fact that binary population synthesis can naturally explain the
observed statistical properties of massive stars (like the observed
ratio of WR to OB stars, relative numbers of various SN types, etc.)
means that single-star models (which represent a minority of stars)
{\it should not} do so.  The point is that by including efficient
rotational mixing and stellar winds that are too strong, single-star
models mimic outcomes that are in fact dominated to a large extent by
binary RLOF.  This raises concerns about the correctness of physical
ingredients adopted in these single-star models, including the
treatment of rotation and turbulent convection in addition to the
overestimated mass-loss rates.

\subsection{Low Metallicity}

The physics of RLOF is governed by the gravitational interaction of
two stars, and is insensitive to the metallicity of the gas being
transferred.  The insensitivity to $Z$ for the extreme mass loss
induced by RLOF is therefore similar, in principle, to the
continuum-driven winds and explosions of LBVs \citep{so06}.  This
should have strong implications for populations of evolved stars and
SNe at low $Z$, although this aspect has not been much explored in the
literature.

Observations do indicate evolution with metallicity, such as the WC/WN
ratio \citep{massey03} and the relative rates of Type Ibc to Type II
SNe \citep{pb03,prieto08,bp09}.  However, we should be cautious that
even binary RLOF may have {\it some} dependence on metallicity, since
metallicity affects the opacity in the star's envelope, and hence, the
hydrostatic stellar radius.  With lower opacity, low-$Z$ stars are
more compact \citep{mm00,heger03}.  The onset of RLOF depends on the
primary star's radius, so a low-$Z$ stellar population might be less
affected by RLOF on average than at $Z_{\odot}$.  It might be easy to
mistakenly attribute such apparent $Z$ dependence entirely to
line-driven winds, so the $Z$ dependence (or not) of RLOF therefore
deserves additional study.  Of course, extrapolating RLOF to low $Z$
also requires detailed knowledge of binary star formation physics and
the resulting period distribution as a function of $Z$, which are not
readily available but may be quite important. This impacts a number of
issues in astrophysics, but most obvious is the progenitors of GRBs.

Nevertheless, even if we were to adopt zero $Z$ dependence for RLOF,
observed trends with $Z$ of WC/WN stars and SNe Ibc/II do not
contradict the dominant role of binary RLOF.  An important point for
interpreting WR subtypes and SN progenitors is that even when
stripping of the H envelope is done by RLOF, the subsequent evolution
from WN to WC (and SN Types IIb to Ib to Ic) is still determined
largely by $Z$-dependent line-driven winds.  This is discussed next.

\section{END RESULTS OF MASS LOSS AND IMPLICATIONS}

\subsection{Outcomes I: WR stars as the product of mass loss}

One of the most fundamental tenets of massive star evolution is that
strong mass loss through winds will strip off a star's H
envelope and leave a bare He core that we observe as a luminous WR
star.  (For the purpose of discussion here, we exclude WNH stars.)
The agent that dominates that stripping of the H envelope and how it
varies with metallicity is a long-standing unsolved issue.  Does every
massive O-type star evolve to become a WR star, or only those in
interacting binaries or in certain initial mass ranges? The answer to
this question has important implications for relative nuclear burning
timescales in various phases, SN progenitors, and many other issues.

In a single-star framework, only the most massive stars are luminous
enough to have radiation-driven winds that can remove the massive H
envelope, so one expects a $Z$-dependent minimum initial mass that can
yield a H-poor WR star, $M_{WR}$($Z$). Standard single-star models
predict $M_{WR} \simeq 35 M_{\odot}$ at $Z_{\odot}$
\citep{heger03,georgy12}, increasing to about 45 and 70 $M_{\odot}$ at
$Z_{LMC}$ and $Z_{SMC}$, respectively \citep{heger03}.  $M_{WR}$ can
be lowered by adopting substantially enhanced mass-loss rates in
models \citep{meynet94,ekstrom12}, or by including the effects of
relatively rapid rotation
\citep{georgy12}.\footnote{\citet{vanbeveren07} has criticized the
  application of these rotating models to observed trends, pointing
  out that the initial 300 km s$^{-1}$ rotation speeds in models are
  not representative of most massive stars.  Correcting observed
  rotation speeds for a distribution of inclination angles, he argues
  that most O-type stars rotate more slowly at 100-120 km s$^{-1}$.}
At first glance, models and observations would seem to be reasonably
well aligned \citep{massey03}: $M_{WR}$ is inferred to be about 25
$M_{\odot}$ at $Z_{\odot}$ \citep{crowther07}, increasing to about 30
and 70 $M_{\odot}$ at $Z_{LMC}$ and $Z_{SMC}$, respectively
\citep{massey00}. However, theoretical expectations for $M_{WR}$ are
based on models that incorporate mass-loss rates that are known to be
a factor of $\sim$3 too high.  (They also do not include the weak-wind
problem discussed earlier, which is important in this mass range, and
which affects a large fraction of SN progenitors.)  With the
$Z^{0.69}$ scaling of mass-loss in line-driven winds \citep{vink01},
this means that the appropriate mass-loss rates for $Z_{\odot}$ are
actually similar to those currently adopted in SMC models.  Grids of
models with appropriate mass-loss rates have not been published, but
we can infer that the net effect will be to move the predicted
$M_{WR}$ upward significantly in single-star models for each $Z$
range, to a point that probably cannot be reconciled with observed
$M_{WR}$.

In a binary evolution paradigm, on the other hand, He stars stripped
of their H envelope can occur over a wider range of initial mass
\citep{vanbeveren07,claeys11}.  In that case, the dominant factors
controlling observationally inferred values of $M_{WR}$ are
detectability and classification, as well as the wind strength that
removes whatever residual H layer may be left at the end of RLOF.
Even if the H envelope is removed (by any mechanism), classifying an
object as a ``WR star'' observationally requires a strong wind to
produce strong emission lines, which will favor sources of higher $L$
and $Z$.  It may be hard to detect exposed He cores resulting from
binary RLOF in $M_{ZAMS}$ = 10-25 $M_{\odot}$, and their low
luminosity and weak emission lines would prevent them from being
classified as WR stars.  Their brighter, cooler companion star may be
overluminous because it has just accreted its companion's H envelope.
Such stripped-envelope stars should be the most common SN~Ibc
progenitors (see below).

Two other key considerations are the observed WC/WN ratio that
increases with $Z$ \citep{massey03}, as well as the fact that even
early-type WR stars in the SMC tend to have some small amount of H
present in their atmospheres \citep{foellmi03}; both of these are
generally atributed to the important role of $Z$-dependent,
line-driven winds.  Again, it is important to recognize the influence
of the stellar wind {\it after} the H envelope is stripped in binary
RLOF.  No matter what mechanism removes the H envelope (binary RLOF,
LBV eruptions, RSG winds, or hot-star winds), the subsequent evolution
will be dominated by a line-driven wind \citep{vanbeveren07,claeys11}.
Binary RLOF leaves a thin H layer, and stars at lower $Z$ or lower $L$
will have a harder time removing it.  Similarly, the evolution from a
WN to a WC star (if this progression is monotonic) will be harder at
lower $Z$ or lower $L$, even if the H envelope was stripped by binary
interaction.  Therefore, comparing models to the observed WC/WN ratio
and showing that it increases with $Z$, for example, is not indicative
of the importance of line-driven winds in any {\it earlier} phases of
evolution, since any scenario should predict an increasing WC/WN ratio
with $Z$.

Altogether, it is difficult to rule out the hypothesis that binary
RLOF is the dominant agent responsible for producing most or all WR
stars, and it remains unclear under what ranges of $L$ and $Z$ (if
any) a single-star can become a WR star via its own wind mass loss.
Unfortunately, finding examples of apparently single WR stars does not
provide a conclusive answer, since a companion star may have exploded
already.  Moreover, the possible importance of RSG mass loss in
producing WR stars is still not well understood.  It is suspicious
that most H-poor WR stars have luminosities of log($L/L_{\odot}$)=5.5
to 5.8 \citep{crowther07,hamann06}, which also corresponds to the
strongest RSG mass loss (Fig.~\ref{fig:mdot}).  The creation of WR
stars is, of course, also closely related to the end fate in a SN and
the rates of various SN subtypes that are seen, as discussed next.

\subsection{Outcomes II:  The main supernova subtypes}

Core-collapse SNe exhibit a wide diversity of properties, summarized
briefly in the accompanying sidebar.  Different SN types are the
direct product of different amounts of mass loss from massive stars,
and they provide key constraints that inform our understanding of
stellar evolution.  Here we discuss the main types of SNe~II and Ibc
with ``normal'' SN atmospheres.  Types~IIn and Ibn with narrow
emission lines from dense CSM that indicate eruptive pre-SN mass-loss
are discussed separately in the next section.  Table 1 includes a map
of SN type to progenitor star properties, based on current prevailing
ideas.

For understanding the main population of SNe, the relative rates of
various subtypes are critical, since they must match the relative
fractions of different progenitor stars for a given initial mass
function (IMF).  Volume-limited rates of the various core-collapse SN
sub-types (excluding SNe~Ia) measured in a controlled sample are now
available \citep{smith11b}; these come from the Lick Observatory SN
Search, which targeted mainly large galaxies representative of
$\sim$$Z_{\odot}$ \citep{li11}.  The most common core-collapse SNe are
Type II-P (48.2$\pm$6\%), marking the explosions of relatively
low-mass RSGs.  Types II-L and IIn contribute 6.4$\pm$3\% and
9$\pm$3\%, respectively.  The remainder (36.5$\pm$6\%) are ``stripped
envelope'' SNe of Types IIb (10.6$\pm$3.6\%) and Ibc (26$\pm$5\%;
including a few peculiar cases like SNe Ic-BL and Ibn that are rare at
$Z_{\odot}$).

The observed fraction of stripped-envelope SNe is a key constraint on
mass loss for the majority of massive stars.  \citet{smith11b} pointed
out that the observed fraction of stripped envelope SNe (IIb+Ib+Ic) of
36.5\% is far too high to be reconciled with predictions of
single-star evolution.  If SN~Ibc progenitors are assumed to be WR
stars, then the observed fraction of SNe~Ibc (not including SNe IIb)
would require $M_{WR}$=22 $M_{\odot}$ \citep{smith11b}.  While this is
not much lower than the lowest-mass WR stars observed in the Milky Way
\citep{crowther07}, it is much lower than can be explained by standard
single-star evolution models --- especially if we recognize that
``standard'' single-star models all adopt mass-loss rates that are too
high.  The observed SN statistics strongly favor the interpretation
that most stripped-envelope SNe (including SNe~IIb) come from
lower-mass stars (10-25 $M_{\odot}$) that lose their H envelope in
binaries.  Again, $\sim$36\% is the observed SN~IIb+Ib+Ic fraction;
compare this to 33\%, which is the fraction of massive stars that
\citet{sana12} expect to have their H envelopes stripped in a binary
system, given the observed binary fraction of O-type stars.  One
infers that binary RLOF can account for the observed
statistics. (Recall that $\sim$80\% of SNe come from initial masses
$<$25 $M_{\odot}$, assuming a Salpeter IMF where every star with
initial mass above 8.5 $M_{\odot}$ explodes as a SN.)  Preference for
the binary channel agrees with relatively low ejecta masses and H/He
mass fractions inferred from detailed radiative transfer models of
stripped-envelope SNe \citep{dessart11,hachinger12,yoon10}, which seem
to rule out the idea that SNe~IIb and Ib can come from progenitors
much more massive than progenitors of SNe II-P, on average.

A dominant binary channel for stripped-envelope SNe is also consistent
with available direct detections and upper limits of SN progenitors.
\citet{smartt09} summarized progress up until 2008, but there have
been several important additions since then.  Observations to date are
consistent with stars that have initial masses of roughly 8-20
$M_{\odot}$ dying as SNe II-P,\footnote{The upper bound of this range
  is uncertain, and higher than found by \citet{smartt09} if one
  accounts for progenitor reddening \citep{we12}.} but this does not
necessarily mean that {\it all} stars in this mass range die that way.
Stars in the same mass range could die as stripped-envelope SNe if
they are in a binary system.  In fact, there are currently 3 direct
progenitor detections for SNe~IIb (SN~1993J, 2011dh, and 2013df) that
are thought to be YSGs in binary systems with inferred initial masses
of 13-17 $M_{\odot}$ \citep{ms09,vandyk13a,vandyk13b}.  Two other SNe
IIb, SN~2001ig and SN~2008ax also show possible indications of a
companion star \citep{ryder06,crockett08}.  Moreover, there are as yet
no detections of progenitors of SNe Ibc.  This seems unlikely if
luminous WR stars are their progenitors \citep{smartt09}, but the hot
temperatures of WR stars make it hard to definitively rule them out.

%------------- IIb - metallicity
What about stripped-envelope SN fractions at lower $Z$?  The SN~Ibc/II
ratio decreases at lower metallicity \citep{pb03,prieto08,bp09}, as
noted earlier.  However, \citet{smith11b} pointed out that the
relevant ratio really is (IIb + Ibc)/(II-P + II-L + IIn), since SNe
IIb are almost identical to SNe Ib but for $<$0.01 $M_{\odot}$ of H in
their outermost layer \citep{dessart11,hachinger12}.\footnote{Many
  studies group SNe IIb and II-L together into a transitional class
  between SNe II-P and Ib in a single-star framework (e.g.,
  Figure~\ref{fig:heger}).  However, it is important to note that SNe
  IIb and II-L are actually quite different, and they are not part of
  the same continuum in decreasing H envelope mass. SNe~IIb really are
  almost identical to SNe~Ib, whereas SNe~II-L have more in common
  with normal SNe~IIn and SNe II-P.}  The recent study by
\citet{arcavi10} finds that compared to giant galaxies at high $Z$,
there is {\it a much larger fraction of SNe~IIb and a lower fraction
  of normal SNe~Ibc in lower-$Z$ dwarf galaxies}.  The lower fraction
of SNe Ibc is expected in single-star models, but the higher fraction
of SNe IIb is not.  Since the removal of the H envelope itself would
be greatly hindered, we would expect a much larger fraction of SNe
II-P and II-L at lower $Z$, not SNe~IIb.  This result is, however,
expected in binary evolution, since low-$Z$ line-driven winds have a
harder time removing the residual H layer that remains after RLOF
\citep{claeys11}.  Interestingly, \citet{arcavi10} also find a larger
relative fraction of SN~Ic-BL in low-$Z$ dwarf galaxies; this is not
yet explained, partly because we don't have a good understanding of
what physical mechanism makes some SNe Ic have such broad lines.  It
follows the trend that GRBs and their associated SNe Ic-BL seem to
prefer low $Z$ \citep{modjaz08}.

Altogether, current evidence strongly suggests that {\it it is no
  longer true that single WR stars are the preferred progenitors of
  most stripped-envelope SNe}.  Massive WR stars might yield {\it
  some} of the SNe~Ibc, of course, especially in the smaller category
of SNe~Ic and GRBs that may favor high-mass progenitors, but {\it many
  SNe Ibc must come from lower initial masses where the H envelope is
  stripped in a binary}.

Using radio and X-ray observations, one can probe the density in the
wind of the SN progenitor directly.  Assuming wind speeds of
$\sim$1000 km s$^{-1}$, observations suggest a very wide range of
mass-loss rates for SN~Ibc progenitors, from $10^{-7} M_{\odot}$
yr$^{-1}$ up to values near the line-driving limit, with an average
around 10$^{-5}$ $M_{\odot}$ yr$^{-1}$ \citep{wellons12}.  The
examples near the lower end of this range are inconsistent with WR
winds, while those near the upper end of the range are (see
Figure~\ref{fig:mdot}).  This is further evidence that SNe~Ibc may
arise from a large range of initial mass, including relatively
low-luminosity stars with weak winds that must have lost their
envelopes in binary RLOF.  Interestingly, the SN~Ibc near the top of
this range tend to show density modulation in their winds, which may
indicate slow\footnote{Note that slower outflows would reduce the
  value of $\dot{M}$ inferred from radio observations.} and dense
outflows from RLOF or interacting winds in binaries
\citep{podsiadlowski92}, pre-SN eruptive mass loss \citep{sa13}, or S
Dorardus (LBV-like) variability \citep{kv06}.  This pre-SN variability
is not yet understood, while much more extreme cases of eruptive
pre-SN mass loss are also seen, discussed next.

\subsection{Outcomes III: Enhanced pre-SN mass loss and luminous SNe}

One of the more exciting new developments in massive star and SN
research in the past decade is the recognition that a subset of
massive stars undergo violent eruptive mass loss immediately
preceeding core collapse, and that this may yield some of the most
luminous SNe in the universe.

A SN blast wave expands outward into the CSM and the ensuing
collision, referred to as ``CSM interaction'', is commonly observed in
the form of X-ray or radio emission \citep{cf94} for normal winds
(Figure~\ref{fig:mdot}).  In 8-9\% of core-collapse SNe
\citep{smith11b}, however, the CSM is so dense that the shock
interaction gives rise to strong narrow emission lines in the
visual-wavelength SN spectrum.  When the CSM is very dense, it can
substantially decelerate the fast SN ejecta and convert a large
fraction of the kinetic energy (10-50\% or more) into radiation.  When
these strong narrow emission lines are observed, we refer to the SN as
Type IIn (narrow H lines) or Ibn (narrow He lines).  In general,
$\dot{M}$ values of at least 10$^{-3}$ to 10$^{-2}$ $M_{\odot}$
yr$^{-1}$ are required for the narrow emission lines to compete with
the luminosity of the normal SN photosphere.  There is a huge
diversity among SNe with strong CSM interaction, which can be
understood in a few different regimes:

\begin{itemize}

%  SLSNe IIn  - very masssive
\item{Super-luminous SNe (SLSNe) with Type IIn spectra represent the
    most extreme cases of eruptive pre-SN mass loss, with luminosities
    $\sim$10 times higher than a normal bright SN Ia.  \citet{sm07}
    proposed that these extremely high luminosities could be achieved
    wth normal energy core-collapse SNe (a few 10$^{51}$ ergs) if the
    fast SN ejecta crash into a very massive 10-20 $M_{\odot}$ CSM
    shell.  This large mass comes from the basic physical requirement
    that the CSM must have enough inertia to substantially decelerate
    the fast SN ejecta and convert its expansion kinetic energy into
    thermal energy that can be radiated away.  Diverse SLSN light
    curve shapes are possible, depending on the distribution of the
    mass and explosion properties
    \citep{vanmarle10,ci11,cw13,moriya13}.  SN~2006gy was the first
    observed event that instigated these ideas of massive CSM shell
    collisions \citep{smith07,smith10b,ofek07,sm07,woosley07}, but a
    number of very luminous SNe IIn have been studied in detail since
    then, including objects like SN~2006tf \citep{smith+08}, SN~2003ma
    \citep{rest11}, and SN~2008fz \citep{drake10}.}

% normal SNe IIn  - massive stars LBVs, eRSGs, YHG
\item{SNe~IIn with moderate luminosity represent less extreme cases
    than SLSNe, but they still require strong CSM interaction that
    indicates eruptive or episodic pre-SN mass-loss events.  Instead
    of pre-SN ejections of 10-20 $M_{\odot}$, more typical
    luminosities require less massive shells of order 0.1-1
    $M_{\odot}$.  The lower luminosity could result from lower
    explosion energy (see SNe IIn-P below), but for normal SNe~IIn it
    is more likely attributed to lower density CSM, or asymmetric CSM
    that only intercepts a portion of the explosion solid angle.  Some
    well studied SNe~IIn like SN1998S \citep{leonard00} and SN~2009ip
    \citep{smith+13,levesque13} are consistent with very asymmetric or
    even disk-like CSM.  The mass-loss rates indicated by the CSM
    suggest that viable progenitors could be LBVs
    \citep{galyam07,taddia13} as well as extreme RSGs or YHGs
    \citep{shr09} (see Fig~\ref{fig:mdot}).  Although the immeditate
    pre-SN eruptive mass loss is less extreme than required for SLSNe
    IIn, some SNe~IIn show strong CSM interaction that continues for
    years or decades (like SN~1988Z, \citealt{art99}) as well as
    long-lasting IR echoes from distant dust shells illuminated by the
    SN \citep{fox11,gerardy02} --- both of these indicate a
    considerable amount of mass lost by the progenitor star for
    centuries before core-collapse, either in previous eruptions or
    very strong dusty winds.  
%% new
    Since the IIn spectral signature depends on mass loss rates and
    not the underlying explosion mechanism, then from the point of
    view of massive-star mass loss, we may naturally expect a
    continuum between normal SNe IIn and either II-L or II-P.
    Identifying this may require larger numbers of SNe to be observed,
    or high-quality early-time spectra.}

% SN IIn-P - low mass 8-10 ecSNe
\item{The subclass of SNe~IIn with plateau light curves, SNe IIn-P,
    was proposed recently \citep{mauerhan13b}, represented by SNe like
    SN~1994W, SN~2009kr, and SN~2011ht.  Among the wide diversity of
    SNe~IIn, this subset is surprisingly homogeneous, with nearly
    identical spectral evolution and very similar light curves that
    all plummet sharply after $\sim$120 days (other SNe~IIn have
    smoothly declining or very slowly declining light curves).  These
    may be relatively low-energy (10$^{50}$ erg) electron-capture SNe
    from $M_{ZAMS}$ = 8-10 $M_{\odot}$ super-AGB stars that achieve
    luminosities of normal SNe through intense CSM interaction
    \citep{smith13b}.}

% Hybrid Type Ia/IIN (or SNe Ia-CSM)
%\item{Some SNe classified as Type IIn may actully be thermocuclear
%    Type Ia SNe that have dense H-rich CSM.  }

%% also H-poor, He-rich SNe Ibn - not related to instability in H envelope
\item{Explosions classified as SNe~Ibn are very similar to SNe IIn,
    except that instead of narrow H lines, they exhibit narrow He
    lines. Their peculiar spectra arise from the same basic scenario
    as SNe IIn, with a SN shock interacting with dense CSM, but here
    the CSM is H-poor.  There are also a few reported transitional
    cases between SNe~Ibn and IIn, including SN~2005la and SN~2011hw
    and 2005la \citep{smith12,pastorello08}, suggesting a possible
    continuum in progenitor H envelope stripping, similar to SNe~II-P,
    IIb, and Ib.  The best studied Type~Ibn is SN~2006jc
    \citep{pastorello07,foley07}, for which an LBV-like outburst was
    detected 2 yr prior to the eventual SN.  If SNe~Ibn come from
    single stars, they are likely to be very massive WR stars;
    however, lower initial masses (10-20 $M_{\odot}$) can yield
    H-depleted progenitors through binary RLOF, and such stars may
    have dense H-poor CSM.  There are not yet any detections of a
    quiescent SN~Ibn progenitor.  Interestingly, \citet{sanders13}
    report the detection of a SN~Ibn in a giant elliptical galaxy with
    no evidence for ongoing star formation at the explosion site,
    which challenges the idea that these arise exclusively from very
    massive progenitor stars.  This may echo the observation that some
    SNe~IIn appear to be caused by an underlying thermonculaer Type Ia
    interacting with dense CSM \citep{silverman13}.}

\end{itemize}
%---------------
% observed progenitors
% observed pre-SN eruptions

Until recently, pre-SN eruptive mass loss and connections to LBVs were
mostly hypothetical, limited to (reasonable) conjectures supported by
the circumstantial evidence that {\it something} must deposit a large
mass of outflowing H-rich CSM so close to the star
\citep{smith+08,smith10b,chugai04}.  However, we now have a handful of
directly detected progenitor stars that seem consistent with LBVs, as
well as a few examples of SN explosions where {\it an outburst} was
actually detected photometrically in the few years before a SN.  In
all cases the SN had narrow emission lines indicative of CSM
interaction, but for each it is also difficult to prove conclusively
that the event was a true core-collapse SN.  So far all these objects
have continued to fade and there is no clear evidence that the stars
survived.

{\bf SN~1961V.}  This is Zwicky's prototype Type V SN, later
associated with LBV-like giant eruptions, but it may have been a true
core-collapse SN~IIn that was not recognized at that time because the
Type IIn class didn't exist yet. A luminous ($-$12.2 mag absolute at
blue wavelengths) progenitor was detected at several epochs for
$\sim$20 yr preceding the SN~1961V event \citep{hds99}, and this
source has now faded.  The pre-SN detections include small ($\sim$0.5
mag) fluctuations that resemble S Dor-like episodes, and in the year
before the SN there is one detection at an absolute magnitude of
roughly $-$14.5, similar to LBV giant eruptions.  Currently, a source
at the same position is about 6 mag fainter than the progenitor, and
still shows narrow H$\alpha$ emission \citep{vdm12} that could
represent ongoing CSM interaction.  The progenitor source has not been
discussed in the context of SN progenitors because the 1961 event was
considered an LBV eruption (a ``super-$\eta$ Car-like event''), not a
true SN (see \citealt{vdm12}).  Recently it was argued that SN~1961V
was actually a true core-collapse SN~IIn \citep{smith11a,kochanek11},
which would provide evidence for a $\sim$100 $M_{\odot}$ LBV-like
progenitor that experienced eruptive mass loss before a SN~IIn.

{\bf SN~2005gl.} This was a normal SN~IIn with pre-explosion {\it HST}
images that showed a source at the SN position, which then faded below
detection limits after the SN had faded \citep{galyam07,gyl09}. Its
high luminosity suggested that the progenitor was a massive LBV
similar to P Cygni, with an initial mass of order 60 $M_{\odot}$ and a
mass-loss rate before core-collapse of $\sim$0.01 $M_{\odot}$
yr$^{-1}$.

{\bf SN 2006jc.} A precursor eruption was discovered in 2004 and noted
as a possible LBV or SN impostor. It had a peak luminosity similar to
that of $\eta$ Car \citep{pastorello07}. No spectra were obtained, but
the coincident SN explosion 2 years later was a Type Ibn with strong
narrow He~{\sc i} emission lines \citep{pastorello07,foley07}.  There
is no detection of the quiescent progenitor.

{\bf SN 2009ip.} This source was initially discovered and studied in
detail as an LBV-like outburst in 2009, before finally exploding as a
much brighter SN in 2012.  A quiescent progenitor star was detected in
archival {\it HST} data, indicating a very massive 50-80 $M_{\odot}$
progenitor \citep{smith10,foley11}. It showed slow variability
consistent with an S Dor LBV-like episode \citep{smith10}, followed by
a series of brief LBV-like giant eruptions
\citep{smith10,mauerhan13,pastorello13}.  SN~2009ip is so far unique
among SN progenitor detections: not only did it have a detection of
the quiescent progenitor and multiple pre-SN eruptions, but unlike any
other object we also have detailed high-quality spectra of the pre-SN
eruptions \citep{smith10,foley11}.  The presumably final SN explosion
of SN~2009ip in 2012 looked like a normal SN~IIn, as the fast ejecta
crashed into the slow material ejected 1-3 years earlier
\citep{mauerhan13,smith+13}. A number of detailed studies of the
bright 2012 transient have now been published, although there has been
some controversy about whether the 2012 event was a core-collapse SN
\citep{mauerhan13,prieto13,ofek13a,smith+13,ofek13a} or some type of
extremely bright non-terminal event
\citep{pastorello13,fraser13,margutti13}.  More recently,
\citet{smith+13} have shown that the object continues to fade and its
late-time emission is consistent with late-time CSM interaction in
normal SNe~IIn.  If SN~2009ip was indeed a SN, it provides the
strongest case that very massive stars above 30 $M_{\odot}$ do in fact
experience core collapse, and LBV-like stars are linked to SNe IIn.

{\bf SN~2010mc.} \citet{ofek13b} reported the detection of a precursor
event $\sim$40 d before the peak of the Type IIn SN 2010mc.  It is
unclear if this was a pre-SN outburst or the SN itself.
\citet{smith+13} showed that the double-peaked light curve of
SN~2010mc was nearly identical to SN~2009ip, for which it has been
suggested that the $-$40 d precursor was actually the faint SN
explosion of a blue supergiant, and the later rise to peak was caused
by CSM interaction.

{\bf SN~2010jl.}  This was a SLSN IIn, with a peak absolute magnitude
brighter than $-$20 mag.  \citet{smith11d} identified a source at the
location of the SN in pre-explosion {\it HST} images, suggesting
either an extremely massive progenitor star or a very young massive
star cluster; in either case it seems likely that the progenitor had
an initial mass above 30 $M_{\odot}$.

%{\bf SN 2011ht.} .... not done, this came out last week....

% -------------------
% implications for evolution

These direct detections of LBV-like progenitors and of pre-SN
outbursts provide unambiguous evidence for violent eruptive mass loss
associated with the latest phases in a massive star's life. This only
occurs in $\sim$9\% of core-collapse SNe \citep{smith11b}. The
extremely short timescale of only a few years probably hints at severe
instability in the final nuclear burning sequences, especially Ne and
O burning \citep{sa13,qs12,am11,sq13}, each of which lasts about 1
yr. These instabilities may be exacerbated in the most massive stars,
athough much theoretical work remains to be done. The increased
instability at very high initial masses is extreme in cases where
pre-SN eruptions result from the pulsational pair instability
\citep{whw02,woosley07}, but eruptions may extend to other nuclear
burning instabilities as well \citep{sa13,am11}. Although the events
listed above are just a few lucky cases, they may also be the tip of
the iceberg. Undoubtedly, continued work on the flood of new transient
discoveries will reveal more of these cases.\footnote{Indeed, as this
  article went to press, \citet{fraser13b} reported the recovery in
  archival data of an outburst 1 yr before the SN~IIn-P 2011ht.} The
limitation will be the existence of high-quality archival data over
long timescales of years before the SNe, but these sorts of archives
are becoming more populated and improved as time passes. When LSST
arrives, it may become routine to detect pre-SN outbursts, although
constraining their physical nature with spectra will still be a
challenge.

Not all SNe with strong CSM interaction necessarily require LBV-like
progenitors, but many of them do, and most of the detected progenitors
or progenitor outbursts are consistent with this hypothesis.  If the
progenitors of SNe~IIn are not actually LBVs, they do a very good
impersonation of the eruption energy, luminosity, spectral morphology,
ejecta mass, H composition, and outflow speeds observed in LBV giant
eruptions.  In any case, the basic observation that very massive stars
are reaching core collapse with massive H envelopes still intact is in
direct conflict with single-star evolution models.  Once again, this
may be related to the issue of overestimated mass-loss rates adopted
in these models and the neglect of binary evolution.

\subsection{Extrapolating to low-Z and Pop III Stars}

Stellar evolution is complicated, and fraught with tremendous
uncertainties about mass loss.  Faced with such road blocks, one can
sympathize with the desire to divert one's attention to Population III
stars, where theorists are unencumbered by observational data.  More
seriously, though, there is indeed great interest in thinking about
what the earliest stars in the Universe might have been like, since
they had a strong impact on their environment and on galaxy evolution
in general.  These stars may have been very massive, and they made the
first SNe, the first stellar mass black holes, the first dust, ejected
the first metals back into the ISM, enhanced or dominated
reionization, and probably triggered the formation of the earliest
generations of low-mass stars \citep{bl04,heger03}.  However, as we
attempt to extrapolate from the huge uncertainty associated with local
stars to a regime where there is no data, we must be cautious.

In studying stars at very low $Z$, a standard approach is to
extrapolate the smooth $Z$ dependence of line-driven winds to very low
or even zero $Z$, essentially assuming that massive stars at the
lowest $Z$ will have no mass loss \citep{heger03}.  This leads to
interesting fates for very massive stars, such as pair instability SNe
\citep{heger03,whw02}.  However, more recent results suggest that
$Z$-dependent winds are weaker than we used to think, even at
$Z_{\odot}$.  On the other hand, binary RLOF and eruptive LBV-like
mass loss are much more important than was appreciated in the past,
and they are relatively insensitive to $Z$.  Thus, it is not safe to
assume that massive stars at very low $Z$ suffer much less mass loss
on average.

As noted above, LBVs are seen in nearby low-$Z$ dwarf galaxies.
Pre-SN mass loss also provides some clues in this regard.
% If the explosive pre-SN mass loss that is seen in Sne IIn is driven
% by some sort of late-phase nuclear burning instability, then there
% is no reason to expect it to be less common at low $Z$.  In fact, m
Many SNe~IIn and SLSNe occur in low-$Z$ dwarf galaxies
\citep{stoll11,neill11}, demonstrating that low $Z$ does not inhibit
strong eruptive mass loss.
% At higher $Z$, the most violent pre-SN eruptive mass loss seems to
% be a phenomenon associated with very massive progenitor stars.  With
% less MS mass loss, low-$Z$ stars should end their lives with a more
% massive core for a given initial mass, and thus, it may be the case
% that a larger fraction of stars are susceptible to these pre-SN
% burning instabilities.
The influence of these pre-SN eruptions may have an interesting effect
on the predicted outcomes of SN explosions at low $Z$
\citep{sa13,co13}.

GRBs, of course, present one of the most puzzling observable
mysteries about mass loss and massive star evolution at low $Z$.  The
fact that GRB progenitors have shed most of their envelopes while
still retaining a great deal of angular momentum \citep{mfw99} would
seem to strongly favor their origin in binary evolution with mass
transfer to spin up the star, or even a merger.  GRBs are quite rare,
so attributing them to a special circumstance like a late-phase merger
is plausible.

\section{SUMMARY AND PERSPECTIVE}

\subsection{Take-home points}

We now have a fairly firm understanding of stellar winds of hot O-type
stars \citep{puls08}, relevant for most of their lives on the
H-burning MS phase.  There remain substantial issues in understanding
the physics of wind driving, magnetic fields, and angular momentum
loss, and how these correspond to various observational diagnostics
--- but in terms of the total mass-loss rates and their impact on
stellar evolution, we know that wind mass loss is weaker than we used
to think, due to the effect of clumping.  The consensus seems to be
that mass-loss rates need to be reduced by a factor of at least 2--3
compared to rates derived observationally from standard $\rho^2$
diagnostics assuming homogeneous winds \citep{dj88,ndj90}.  Reduction
by a factor of $\sim$10 is probably an overstimate for most O-type
stars, but not for later O-types and early B-type MS stars subject to
the weak-wind problem.

Astronomers often scoff at a factor of 2 in any individual
measurement, but we must recognize that systematically lowering
$\dot{M}$ by even a factor of 2 in models is very significant, and a
factor of 3 is huge (current debate generally centers around a factor
of 2 or 3).  A factor of only 2 lower $\dot{M}$ would be like
replacing mass-loss rates at $Z_{\odot}$ with mass-loss rates
currently used for 0.37 $Z_{\odot}$ (i.e. lower than the LMC), whereas
a factor of 3 would correspond to 0.2 $Z_{\odot}$, similar to models
for SMC stars (using $\dot{M} \propto Z^{0.69}$,\citealt{vink01}).
With moderatly weaker mass loss, we know that SMC models exhibit
significantly different evolution than Milky Way stars.  Most
importantly, they do not produce such a large population of H-free WR
stars.  However, it is over this same metallicity range
\citep{massey03} that we are testing the validity of the physical
ingredients in stellar evolution models.  This isn't good.

For post-MS supergiant phases and binary RLOF, the uncertainty in mass
loss is far worse --- but these are also times when mass loss is
strongest!  Given that single-star models neglect eruptive mass loss
and binaries, but match many properties of the observed distributions
anyway, predictions beyond the end of H burning are not unique.
Extrapolations to low $Z$ and prescriptions for feedback from early
stellar populations are therefore highly uncertain.  Researchers
working in other fields where massive star feedback is relevant ---
galaxy evolution, reionization, chemical evolution, low-metallicity
stars, etc. --- should be cognizant of this.

\subsection{Perspectives and Directions}

With such a high level of uncertainty in the most important phases of
mass loss for massive stars, astronomers working on mass loss and
massive star evolution have a few choices on how to proceed: 1)
restrict their attention to MS stars, 2) work on Population III stars
(see above), or 3) confront the complicated effects of binaries and
time-dependent mass loss, as well as their influence on observed
populations.

A major hurdle is disentangling the observed effects of single-star
mass loss and binary RLOF.  On the one hand, Occam's razor encourages
us to keep things simple, motivating us to understand single stars
before we can hope to tackle the ``free parameter heaven'' of
binaries.  On the other hand, the massive stars we observe are mostly
binaries: A wise man once said that we should ``Make things as simple
as possible, but not simpler''.  The message here is that ignoring
binaries, but testing single-star models against observed statistical
properties of massive stars (which are mostly binaries) leads us to
misinterpret single stars too, because we are relying upon strong
winds and rotation to compensate for actual outcomes of binary RLOF.

% -------------

Since the observed statistics of massive-star populations are
unavoidably contaminated by the effects of binary stars, one
alternative approach for testing single-star models would be to place
less emphasis on matching these observed {\it statistics}, and more
emphasis on matching individual observed stars with well-constrained
physical parameters.  Masses for ``effectively single'' stars can be
measured in wide binary systems where the stars will not interact (or
O-type stars that have not yet interacted on the MS), and for these a
detailed modern quantitative analysis can yield estimates of
$\dot{M}$, $T_{eff}$, $L$, rotational speed, abundances, etc.  The
distance must be known, but the Magellanic Clouds and clusters with
known distances can help, and {\it GAIA} will soon lower distance
uncertainties for the nearest massive stars in the Milky Way.  Doing
this for a number of stars at various evolutionary stages and
luminosities will build up a ``grid'' of observational constraints
that models can aim to fit.  Much of this observational work has
already been done.

Given the huge level of uncertainty in post-MS mass loss (for both
RSGs and LBVs), another general approach for theorists could be to use
stellar evolution codes as ``toy models'', to investigate the final
outcome for a wide range of possible mass-loss prescriptions that
include episodic mass loss.  For example, one could calculate
evolutionary tracks for very massive stars that have lower wind
$\dot{M}$ during H burning, and then simply invoke one, two, or
multiple sudden ejections of 10 $M_{\odot}$ in post-MS phases to mimic
LBV eruptions, in order to learn how the star's further evolution
responds.  This approach may seem artificial, but the goal would be to
reverse-engineer stars, not to provide uniquely correct evolutionary
tracks.  In any case, much additional work on constraining post-MS
mass loss of all forms is needed.  A number of specific possible
future directions for theory and observations are listed below.

\section{FUTURE ISSUES: For theorists}

1.  A major task is to calculate stellar evolution models with lower
wind mass-loss rates for MS phases, including rates appropriate for
the weak-wind problem in later O-type and early B-type stars.  These
can be compared to precise measurements of physical parameters for
well-studied individual stars, rather than populations, as noted
above.  For post-MS phases, mass-loss is simply too uncertain to
predict unique outcomes.  Toy models with various prescriptions for
unsteady $\dot{M}$ can help constrain the possible outcomes and their
influence on SNe.  Perhaps new open-source codes like MESA
\citep{paxton11} will facilitate this.

2.  For episodic mass-loss in the latest pre-SN evolutionary phases,
further work on stellar evolution with hydrodynamics and advanced
nuclear burning is essential.  Pre-SN eruptions constitute a new and
very important unsolved problem in astrophysics, which may influence
the outcome of core-collapse itself.

3.  Theoretical investigations are needed to probe the underlying
cause of LBV eruptions and its impact on the subsequent stellar
evolution.  Predictions for observed consequences of envelope
instabilities, mergers/collisions, and explosive burning events are
needed to link various possible physical mechanisms to observables for
transient sources.

4.  The hydrodynamics of RLOF and its dependence (or not) on
metallicity is an important issue.  A better understanding of mass and
angular momentum transfer vs. mass loss from binary systems is needed
to determine what conditions lead to mergers.  In principle, mass
transfer itself should be insensitive to $Z$, but in practice, $Z$
affects opacity, the stellar radius, and thus the onset of RLOF.  In
the end we may find that observed trends with $Z$ hinge upon the
formation of massive binaries \citep{kratter10} and its dependence on
fragmentation and cooling as a function of $Z$.

5.  We need additional theoretical investigations of RSG mass loss
and its scaling with initial mass, $Z$, and as function of
evolutionary time.  How does the global $\dot{M}$ depend on the
interplay between pulsational instability and dust formation?

\section{FUTURE ISSUES: For observers}

1.  To help constrain single-star evolution, we need precise estimates
of physical parameters for a number of nearby ``effectively single''
stars at various evolutionary stages that can serve as anchors for
stellar evolution models.  Relatively wide binaries that have not yet
interacted are good targets for this.

2.  To better understand how binarity influences observed trends,
continued work on the binary fraction and orbital parameters as a
function of environment: clusters vs. field, and $Z$.  Orbits are
often assumed to be circular in models, but the eccentricity
distribution may also be critical for some problems.

3.  We need better constraints on the episodic mass loss of LBVs,
including the lifetime of the LBV phase, the total mass lost per
eruption, the duty cycle of eruptions, and how these vary with $Z$ and
the initial masses and binarity of central stars.  The large number of
shell nebulae recently discovered by IR surveys
\citep{wachter10,gvaramadze10} may be very helpful in improving
statistics and associations with the central stars.  This is needed to
guide prescriptions for including episodic mass loss in models.

4.  Regarding observational constraints on the total mass lost in
binary RLOF as compared to mass-transfer rates, studies of RLOF
binaries or post-RLOF systems can help constrain under what conditions
mass transfer is conservative, or to what degree it is.

5.  Ongoing studies of RSGs in clusters may help provide better
constraints on measured RSG $\dot{M}$ as a function of evolutionary
stage for a range of $Z$.  For how long and for which stars is the
strongest mass loss (self-obscured, masers) at work?  Connections to
pulsation amplitudes and stars on blueward tracks are also important.

6.  SN progenitors and SNe with CSM provide critical clues about the
mass loss of massive stars, and well constrained individual cases are
limited to relatively nearby SNe. This provides important links
between stars and their end fates, and has opened a new window for
observing episodic mass loss in the very final stages of evolution.
Also, among nearby stars, we need better constraints on the properties
of the faint He-rich stars in binaries that have lost their H envelope
to a companion in RLOF, because they may be the dominant stripped
envelope SN progenitors.

7.  Stellar evolution codes adopt time-averaged mass-loss
prescriptions, but increasing evidence suggests that brief disruptive
events are very important.  Observational work on transient phenomena
in general is still a relatively new topic, providing important
constraints on physical parameters and rates for these events.  Are
there observational signatures that will tell us if they are binaries?

8.  Studies of SN and transient environments, including their
surrounding stellar populations, can help disentangle the relative
evolutionary phases of their progenitors (ages, burning stages, etc).
This may help to idenfy types of stars that don't fit with the
expected age of their surrounding populations, perhaps flagging stars
that are mainly binaries.

\section{ACKNOWLEDGEMENTS}

I am grateful for numerous discussions with many people, but
especially Stan Owocki, Jo Puls, Dave Arnett, Paul Crowther, Jorick
Vink, Jose Prieto, Todd Thompson, Norbert Langer, Selma de Mink, Luc
Dessart, John Hillier, Ben Davies, John Eldrige, Phil Podsiadlowski,
and Matteo Contiello.

\smallskip

{\bf This is the end of the main article text.  The following sections
  include a glossary of terms and sidebars.}

\section{GLOSSARY}

SN: Supernova.  For massive stars this usually corresponds to the
collapse of the iron core, but a SN could also result from an
electron-capture SN (8--10 $M_{\odot}$) or pair instability explosion.

Metallicity ($Z$).  Relative abundance of elements heavier than He.
Plays an important role for radiatively driven winds and for the
opacity in parts of a star's envelope.

MS: Main Sequence.  Core-H burning lifetime of massive stars, when
they appear as O-type stars or early B-type stars.  Some WNH stars are
also MS stars.

ZAMS: Zero-Age Main Sequence.

Massive star: defined here as initial mass $M_{ZAMS}$ above 8
$M_{\odot}$, which experiences a core collapse or other SN at death.
Note that some stars below this limit initially might gain mass in a
binary system and explode as a SN.

non-LTE: Radiative transfer calculations not invoking local
thermodynamic equilibrium.

P Cygni profile.  Blueshifted absorption from an outflowing wind.

Eddington limit ($\Gamma$=1).  Point where a star's luminosity is so
strong that radiation force balances gravity.  One expects such a
situation to be accompaied by severe mass loss.

RLOF: Roche lobe overflow.  A phase in the evolution of binary systems
when either star has a radius that exceeds the inner Lagrane point
(L1) and begins to transfer mass to the other star.

CSM: circumstellar material, or circumstellar medium.

ecSN: electron-capture SN.  Collapse of degenerate ONeMg core due to
electron capture in lower-mass (8-10 $M_{\odot}$) SN progenitors.

PPI: pulsational pair instability.

PISN: Pair instability SN.

\section{SIDEBAR: Types of H-burning massive stars (Sec 2)}

OB dwarfs: Hot massive main-sequence stars with luminosity class V,
early in core-H burning.  {\it Examples: $\theta^1$ Ori C, $\zeta$
  Oph}

Of, Oe, Ofpe/WN9 (etc) supergiants: O stars with evidence for strong
winds in their spectra (i.e. emission lines).  These are thought to be
more evolved than O dwarfs, either in the later phases of core-H
burning, or in transition to He burning.  {\it Examples:} $\zeta$
Puppis, Hen 3-519, S61.

WNH: or more colloquially, ``O-stars on steriods''.  These are H-rich
stars with WR signatures in their spectra.  If single, these stars are
in the late stages of core-H burning MS evolution of the most massive
stars.  If in binary systems, these are the possible products of
mergers, like more massive analogs of blue stragglers.  These are the
most massive stars in young massive star clusters, and the most
massive stars measured in binaries.  {\it Examples:} WR25, WR20a.

\section{SIDEBAR:   Evolved Massive Stars I: Cool Types (Sec 3)}

RSG: Red supergiant.  Coolest evolved massive stars. Strongest
mass-loss phase for single stars with initial masses below about 30
$M_{\odot}$. {\it Examples:} Betelgeuse.

Extreme RSG (OH/IR star): The most luminous RSGs whose mass loss is so
strong that they are self-obscured at visual wavelengths, with strong
IR excess from dust and maser emission from OH, SiO, and H$_2$O.  {\it
  Examples:} VY~CMa, NML Cygni.

YSG: Yellow supergiant. Rare stars that appear in the middle of the HR
diagram, possibly as post-RSGs, and usually with strong mass loss
\citep{drout12}.  This is a short-lived phase in both binary and
single-star models. {\it Examples:} progenitor of SN~2011dh.

YHG: Yellow hypergiant. The most luminous YSGs, usually with extreme
mass loss.  Often designated with a luminosity class of Ia$^+$
\citep{dj98}.  {\it Examples:} IRC+10420, HR~5171A, $\rho$ Cas.

\section{SIDEBAR:   Evolved Massive Stars II: Hot Types (Sec 4)}

WR.  Wolf-Rayet star.  He burning massive stars with very strong
emission lines of He in their spectra, caused by very strong winds.
WN (WR with N) and WC (with C lines) are exposed He cores of massive
stars that have lost their H envelopes through prior mass loss. {\it
  Examples: $\gamma^2$ Vel, EZ CMa}

LBV: Luminous blue variable.  A group of evolved massive stars that
exhibit eruptive mass loss or irregular variability.  A union of
various subtypes, including giant eruptions ($\eta$ Car variables), S
Dor variables, $\alpha$ Cyg variables, P Cygni stars, Hubble-Sandage
variables, etc.  Most have strong winds and strong emission-line
spectra.  Candidate LBVs are stars which have similar spectra and/or
dust shells, but have not yet exhibited variability.  {\it Examples:}
$\eta$ Car, P Cygni, AG Car, S Dor, HR Car.

BSG: Blue supergiant.  Post-MS sequence massive stars with B spectral
types.  The relative number of BSGs in observed HR diagrams of stellar
populatios is not well understood.  {\it Examples:} Sk $-$69~202, Sher
25.

Be and B[e] Stars:  B-type stars with strong and usually time-variable
emission lines, often showing evidence for disk-like CSM.  Be stars
are rapid rotators, possibly resulting from increased angular momentum
through mass accretion in binaries.  The B[e] stars have strong
forbidden line emission and IR excess from dust, thought to arise in a
circumstellar disk or torus.  Some are high-luminosity evolved
supergiants similar to LBVs.  {\it Examples:} $\gamma$ Cas (Be), R4 in
the SMC (B[e]).

\section{SIDEBAR: SN Subtypes (Sect 6)}

SN explosions exhibit a wide diversity of observed properties.
Proliferating classifications are based on their spectra and light
curves, not a physical mechanism.  Types I and II are determined by
the presence (II) or absence (I) of H lines in the spectrum. Main
categories of SNe are listed here (see also \citealt{filip97}).

{\bf Type II-P.}  These are the most common type of core-collapse SN,
exhibiting broad H lines in the spectrum, and showing a plateau of
typically $\sim$100 days in the visual-wavelength light curve.  These
arise primarily from RSGs with initial masses of 8--20 $M_{\odot}$
\citep{smartt09} and mass-loss rates of 10$^{-6}$-10$^{-5}$
$M_{\odot}$ yr$^{-1}$.

{\bf Type II-L.}  Spectroscopically these are very similar to SNe
II-P, but their light curves show a linear decay. The faster decline
probably results from a lower-mass H envelope, indicating heavier
pre-SN mass loss by the progenitor.

{\bf Type II-pec.}  Similar to SNe~II-P spectroscopically, but with a
light curve that rises slowly from an initially faint state due to a
more compact BSG progenitor, with SN~1987A being the prototype.

{\bf Type IIn.}  SNe with prominent narrow H lines in their spectra.
The narrow lines arise in nearby (a few 10$^{15}$ cm) CSM that is
photoionized or shock-heated by the SN. These SNe require strong mass
loss immediately (a few years) preceding the SN and exhibit wide
diversity.

{\bf Type IIn-P.}  A new subclass of SNe IIn with plateau-shaped light
curves, which show a pronounced (3-6 mag) drop in flux at times around
120 days \citep{mauerhan13b}.
% \citet{smith13b} suggested that this type of SN made the Crab
% Nebula.

{\bf Type IIb.}  These are very similar to Type Ib, except that they
show transient broad H lines in their early-time spectra.  This is due
to a low mass (0.01 $M_{\odot}$ or less) residual H envelope remaining
on the outer layers of the progenitor.

{\bf Type Ib.}  SNe which do not show H in their spectra, but which
have strong broad He lines.

{\bf Type Ibn.}  Analogous to SNe~IIn caused by strong CSM
interaction, but with strong narrow lines of He instead of H.

{\bf Type Ic.}  SNe which show no H, and little or no He lines in
their spectra, requiring the most extreme examples of
stripped-envelope progenitors.
 
{\bf Type Ic-BL/GRB.}  These are SNe~Ic with very broad (20,000-30,000
km s$^{-1}$) lines in their spectra.  This is the only type of SN
observed to be associated with GRB explosions.

{\bf SLSNe.} In principle, this class would include any of the above
with a peak absolute magnitude more luminous than about $-$20 or $-$21
mag (i.e. brighter than the brightest SNe~Ia).  In practice, only
spectral types IIn, II-L, and Ic have been seen in this class so far.

{\bf Type Ia.}  Thermonuclear SNe from white dwarf progenitor stars.

{\bf Type Ia/IIn (Ia-CSM).}  SNe Ia with narrow H lines in their
spectra, indicating dense CSM.  When strong CSM interaction veils
underlying Ia spectral features, it is difficult to distinguish these
from core-collapse SNe~IIn.

\section{SIDEBAR: Super-luminous SNe without narrow lines (Sect 6.3)}
% SLSNe
The most luminous SNe seen so far are of Types Ic or II-L, without
narrow lines.  These SLSNe Ic \citep{quimby11,galyam12} and II-L
\citep{miller09,gezari09} might be explained by CSM interaction
with a dense shell in special cases where the shell has relatively
sharp outer boundary. After shock breakout from the wind, radiation
diffuses out from the accelerated CSM, yielding no narrow lines
\citep{sm07,ci11}.  However, these SNe may also be explained by
magnetar birth \citep{kb10,woosley10}, so they are not necessarily the
product of eruptive pre-SN mass loss.  SLSNe Ic do, of course, require
a stripped H envelope like other SNe Ic, and they are interestingly
similar to SNe Ic-BL and GRBs in that they seem to prefer low-Z host
galaxies \citep{neill11}.  This may also suggest a binary origin for
these stars.

\begin{table}\begin{center}%\begin{minipage}%1
\def~{\hphantom{0}}
\caption{Mapping of SN types to their likely progenitor star properties}
\label{tab1}
\begin{tabular}{@{}llccc@{}}
  \toprule
  SN       &Progenitor Star$^a$  &$M_{ZAMS}$     &$\dot{M}$$^b$ &$V_{\infty}$     \\
  ...      &...         &($M_{\odot}$)  &($M_{\odot}$ yr$^{-1}$) &(km s$^{-1}$)   \\ 
  \colrule
  II-P     &RSG            &8--20         &10$^{-6}$--10$^{-5}$    &10-20        \\
  II-L     &RSG/YSG        &20--30 (?)    &10$^{-5}$--10$^{-4}$    &20-40        \\
  II-pec   &BSG (b)        &15--25        &10$^{-6}$--10$^{-4}$    &100-300        \\
  IIb      &YSG (b)        &10--25        &10$^{-5}$--10$^{-4}$    &20-100       \\
  Ib       &He star (b)    &15--25 (?)    &10$^{-7}$--10$^{-4}$    &100-1000     \\
  Ic       &He star (b)/WR &25--?         &10$^{-7}$--10$^{-4}$    &1000         \\
  Ic-BL    &He star (b)/WR &25--?         &10$^{-6}$--10$^{-5}$    &1000         \\
  \hline
  IIn (SL) &LBV            &30--?         &(1--10)                &50-600      \\
  IIn      &LBV/B[e] (b)   &25--?         &(0.01-1)               &50-600      \\
  IIn      &RSG/YHG        &25--40        &10$^{-4}$--10$^{-3}$    &30-100       \\
  IIn-P    &super-AGB          &8--10         &0.01-1                 &10-600       \\
  Ibn      &WR/LBV         &40--?         &10$^{-3}$--0.1          &1000        \\
  \hline
  Ia/IIn   &WD (b)         &5-8 (?)       &0.01-1                 &50-100      \\
  \botrule
\end{tabular}\end{center}%\end{minipage}
{$^a$Most likely progenitor star type. ``(b)'' indicates that a binary channel is probably key.  Note that stars which shed envelopes in binary RLOF are likely to have a slow (10 km s$^{-1}$) equatorial outflow, in addition to the wind speed of the star. \\ $^b$Mass-loss rates for pre-SN eruptions are listed in parentheses, corresponding roughly to the total mass ejected in the few years immediately preceding core-collapse. The mass-loss rates may be lower but still substantial at larger radii traced by the expanding SN shock at late times.}
\end{table}

\end{document}